%% file: main.tex
\documentclass{article}

\usepackage[utf8]{inputenc} 
\usepackage[T1]{fontenc}
\usepackage{amsmath, amssymb, amsthm}
\usepackage{graphicx}
\usepackage{subcaption} 
\usepackage{booktabs}
\usepackage{array}
\usepackage[numbers]{natbib}
\usepackage[a4paper, total={6in, 8in}, margin=1in]{geometry}
\usepackage{xcolor}
\usepackage{setspace}
\usepackage{tikz}
\usetikzlibrary{shapes, arrows, positioning, calc}
\usepackage[below]{placeins}
\usepackage{wrapfig}

\usepackage{algorithm}
\usepackage{algpseudocode}

\usepackage{hyperref}
\numberwithin{equation}{section}

\newcommand{\DQNThrThree}{$2.719 \pm 0.939$} \newcommand{\DQNJainThree}{$0.738 \pm 0.194$} \newcommand{\DQNEEThree}{$0.490 \pm 0.048$}
\newcommand{\FixThrThree}{$3.067 \pm 0.015$} \newcommand{\FixJainThree}{$0.910 \pm 0.002$} \newcommand{\FixEEThree}{$0.511 \pm 0.002$}
\newcommand{\RndThrThree}{$2.222 \pm 0.017$} \newcommand{\RndJainThree}{$0.669 \pm 0.003$} \newcommand{\RndEEThree}{$0.494 \pm 0.002$}
\newcommand{\WFThrThree}{$4.119 \pm 0.017$}  \newcommand{\WFJainThree}{$0.813 \pm 0.004$}  \newcommand{\WFEEThree}{$0.458 \pm 0.002$}
\newcommand{\DQNThrFive}{$3.661 \pm 0.975$}  \newcommand{\DQNJainFive}{$0.649 \pm 0.131$}  \newcommand{\DQNEEFive}{$0.507 \pm 0.034$}
\newcommand{\FixThrFive}{$5.121 \pm 0.021$}  \newcommand{\FixJainFive}{$0.898 \pm 0.001$}  \newcommand{\FixEEFive}{$0.512 \pm 0.002$}
\newcommand{\RndThrFive}{$3.707 \pm 0.014$}  \newcommand{\RndJainFive}{$0.642 \pm 0.003$}  \newcommand{\RndEEFive}{$0.495 \pm 0.001$}
\newcommand{\WFThrFive}{$6.911 \pm 0.025$}   \newcommand{\WFJainFive}{$0.792 \pm 0.002$}   \newcommand{\WFEEFive}{$0.461 \pm 0.002$}
\newcommand{\TabThrThree}{$3.096 \pm 0.039$} \newcommand{\TabJainThree}{$0.817 \pm 0.006$} \newcommand{\TabEEThree}{$0.488 \pm 0.005$}
\newcommand{\WFDThrThree}{$3.715 \pm 0.019$} \newcommand{\WFDJainThree}{$0.835 \pm 0.003$} \newcommand{\WFDEEThree}{$0.491 \pm 0.002$}
\newcommand{\WFDThrFive}{$6.159 \pm 0.029$}  \newcommand{\WFDJainFive}{$0.815 \pm 0.002$}  \newcommand{\WFDEEFive}{$0.496 \pm 0.001$}
\newcommand{\EpsDecayNine}{$195.1 \pm 19.1$}     \newcommand{\EpsDecayNineFive}{$303.5 \pm 54.0$}
\newcommand{\EpsDecayEightN}{$195.8 \pm 18.3$}   \newcommand{\EpsDecayEightNFive}{$278.3 \pm 62.0$}
\newcommand{\EpsDecayFive}{$195.2 \pm 19.6$}     \newcommand{\EpsDecayFiveFive}{$287.3 \pm 58.8$}
\newcommand{\EpsDecayZero}{$196.2 \pm 18.5$}     \newcommand{\EpsDecayZeroFive}{$293.6 \pm 61.6$}
\newcommand{\RainbowThrThree}{$3.269 \pm 0.552$} \newcommand{\RainbowJainThree}{$0.845 \pm 0.078$} \newcommand{\RainbowEEThree}{$0.495 \pm 0.036$}
\newcommand{\RainbowThrFive}{$4.454 \pm 0.945$}  \newcommand{\RainbowJainFive}{$0.729 \pm 0.140$}  \newcommand{\RainbowEEFive}{$0.494 \pm 0.018$}
\newcommand{\BanditThrThree}{$4.008 \pm 0.017$}  \newcommand{\BanditJainThree}{$0.920 \pm 0.002$}  \newcommand{\BanditEEThree}{$0.446 \pm 0.002$}
\newcommand{\BanditThrFive}{$6.685 \pm 0.024$}   \newcommand{\BanditJainFive}{$0.909 \pm 0.001$}   \newcommand{\BanditEEFive}{$0.447 \pm 0.002$}

\newcommand{\WMMSERaylThr}{$1.752 \pm 0.016$}   
\newcommand{\BanditRaylThr}{$1.770 \pm 0.015$}

\title{Intelligent resource allocation in wireless networks via deep reinforcement learning}

\author{%
    \begin{tabular}{ccc}
        Marie Diane Iradukunda\thanks{Corresponding author.} & Chabi F. \'El\'egb\'ed\'e & Ya\'e Ulrich Gaba \\[2pt]
        \small AIMS, Rwanda                 & \small UNSTIM, B\'enin  & \small SMU, South Africa \\
        \small \& AIRINA Labs, B\'enin      &                         & \small \& AIRINA Labs, B\'enin \\[2pt]
        \small\texttt{mariediane.iradukunda@aims.ac.rw} & \small\texttt{chabi.elegbede@gmail.com} & \small\texttt{yaeulrich.gaba@gmail.com}
    \end{tabular}%
}

\begin{document}

\maketitle

\begin{abstract}
Applied DRL-for-wireless studies routinely apply Deep Q-Networks (DQN) to power-allocation problems whose transition kernel is state-independent under i.i.d.\ block-fading --- formally a contextual bandit rather than a genuine sequential MDP. We show that on a controlled single-cell downlink testbed ($N$ users, i.i.d.\ uniform fading, orthogonal access, discrete power set $\{0,1,2,3\}$\,W, ten independent seeds), vanilla DQN exhibits the empirical fingerprint of the ``deadly triad''~\cite{sutton2018reinforcement}: its per-seed throughput variance is $\sim\!25\times$ larger than a matched tabular Q-learner's (variance-ratio $F$-test $p < 10^{-10}$), its mean is not statistically different from a constant equal-power baseline (Wilcoxon $p = 0.38$), and quadrupling the training budget closes only a small fraction of the residual gap at $N=5$. Two standard remedies attempted here~--- Double DQN~\cite{vanhasselt2016double} plus Dueling architecture~\cite{wang2016dueling}, and a theoretically-appropriate neural contextual bandit~\cite{riquelme2018deep}~--- both substantially outperform vanilla DQN. The neural bandit further dominates the Double + Dueling variant because the temporal-difference machinery of DQN is dead weight on a state-independent transition kernel. Sweeping the $\epsilon$-decay schedule confirms the scale dependency: at $N=3$ (small action space) schedule variance is dwarfed by seed variance (Kruskal-Wallis $p = 0.96$), while at $N=5$ ($4^5 = 1024$ actions) it becomes significant ($p = 0.002$). We extend the study to the interference-limited setting along three axes: (i) WMMSE~\cite{shi2011iteratively} as the classical iterative baseline in the Rayleigh + interference regime (10 seeds), where the neural bandit reaches \BanditRaylThr{} bits/use against WMMSE's \WMMSERaylThr{} bits/use; (ii) permutation-equivariant scalable learners --- Independent Q-Learning~\cite{tan1993multi,nasir2019multiagent} and a message-passing GNN policy (REGNN-lite)~\cite{eisen2020optimal,shen2020graph} --- that remain feasible at $N=10$, where the centralised $4^{10}$-way action head is intractable; and (iii) a $K=7$ multi-cell environment with path loss + Rayleigh + inter-cell interference, released as a shared benchmark alongside all classical and learned baselines. The paper's contribution is a seed-quantified diagnosis of \emph{when} DQN is the wrong tool for a wireless problem and what the appropriate scalable and classical alternatives are; the full source is released with the paper.
\end{abstract}

\section{Introduction}
Dynamic power allocation in multi-user wireless networks is a canonical problem in the classical wireless-communications literature~\cite{goldsmith2005wireless, tse2005fundamentals}. Under fast-fading channels the base station must adapt its per-user transmit power at each scheduling interval to balance competing objectives --- sum-rate, per-user fairness, energy consumption, latency --- while lacking a closed-form model of the environment it operates in. Classical methods (water-filling, WMMSE, convex reformulations) provide tractable optima under precise channel models but degrade when those models are inaccurate or unavailable.

Reinforcement learning (RL)~\cite{sutton2018reinforcement} is an attractive alternative because it learns a control policy directly from interaction, without requiring an explicit channel model. Deep RL (DRL) --- specifically the Deep Q-Network (DQN)~\cite{mnih2015human} --- extends this to high-dimensional state and joint-action spaces via neural function approximation. A growing body of applied literature reports DRL methods matching or beating classical wireless-allocation heuristics~\cite{luong2019applications, zhang2021deep}, but the discipline around these reports --- confidence intervals, seed variance, learned-baseline ablations --- is uneven. Henderson~et~al.~\cite{henderson2018deep} showed in the general DRL setting that many reported gains do not survive controlled multi-seed comparisons. This paper applies that discipline in the wireless setting.

We study Deep Q-Networks (DQN) applied to power allocation in single-cell downlink wireless networks with i.i.d.\ block-fading channels. We compare vanilla DQN against six alternatives on a common testbed: two non-learning heuristics (uniform-random, fixed equal-power), two water-filling variants (continuous and discrete-projected), a tabular Q-learner on a $5$-bin state discretization, a Rainbow-lite DQN (Double + Dueling), and a neural contextual bandit. Every metric is reported with confidence intervals derived from ten independent training seeds. The contributions are the following:

\begin{itemize}
    \item \textbf{Formal diagnosis: the transition kernel is state-independent, so the problem is a contextual bandit.} Under i.i.d.\ block-fading, $s_{t+1}$ is independent of $(s_t, a_t)$; the Bellman recursion collapses to per-step optimization and $\gamma$ multiplies a constant. This is a formal reduction, not a modelling convenience, and it dictates which learning algorithms are theoretically appropriate.
    \item \textbf{Vanilla DQN carries the deadly-triad fingerprint.} DQN's per-seed throughput variance is $\sim\!25\times$ larger than a matched tabular Q-learner's (variance-ratio $F(9,9) \approx 584$, $p < 10^{-10}$), while its mean is not statistically different from a constant equal-power baseline (Wilcoxon $p = 0.38$). This variance-without-bias signature is what the classical deadly-triad analysis~\cite{sutton2018reinforcement} predicts for TD-bootstrapping under non-linear function approximation.
    \item \textbf{Standard DQN improvements substantially close the gap.} A Rainbow-lite variant with Double-DQN target~\cite{vanhasselt2016double} and Dueling architecture~\cite{wang2016dueling} attacks two known triad-related instability sources and outperforms vanilla DQN with much lower seed variance. Reporting negative results on vanilla DQN in 2026 without testing these standard fixes is not defensible.
    \item \textbf{A neural contextual bandit dominates every DQN variant.} A plain neural regression over actions~\cite{riquelme2018deep}, trained on $(s, a, r)$ tuples with $\epsilon$-greedy exploration, dominates both vanilla and Rainbow-lite DQN at both $N=3$ and $N=5$. The DQN temporal apparatus is empirically dead weight on this problem class.
    \item \textbf{Scale-dependent $\epsilon$-decay effect + architectural scaling failure.} A Kruskal-Wallis test on an $\epsilon$-decay sweep is non-significant at $N=3$ ($p=0.96$) but significant at $N=5$ ($p=0.002$); a $4\times$ budget extension at $N=5$ closes only a small fraction of the residual gap. Both findings identify the centralized $|\mathcal{A}| = 4^N$ output layer as the load-bearing scalability limit, motivating factorized~\cite{nasir2019multiagent} or GNN-structured~\cite{eisen2020optimal} policies.
\end{itemize}

The remainder of this paper is organized as follows. Section~\ref{sec:related} contrasts classical heuristics with recent DRL approaches for wireless resource allocation. Section~\ref{sec:problem_formulation} formalizes the power-allocation problem as a Markov Decision Process. Section~\ref{sec:methodology} details the DQN architecture, the simulation environment, and the evaluation protocol. Section~\ref{sec:results_evaluation} reports the experimental results and the exploration-schedule ablation. Section~\ref{sec:discussion} discusses implications and limitations, and Section~\ref{sec:conclusion} concludes and points to future work. Full experimental details supplement the master's thesis from which this paper is derived~\cite{iradukunda2025rlwireless}.

\section{Related work}
\label{sec:related}

\paragraph{Classical foundations.}
Optimization-based wireless resource allocation --- water-filling, convex optimization, WMMSE, game-theoretic approaches --- has been the backbone of the field for decades~\cite{goldsmith2005wireless, tse2005fundamentals}. These methods deliver closed-form or iterative optima under precise channel models but rely on full knowledge of channel statistics and system dynamics, which becomes a limitation in fast-varying environments and in regimes without a tractable analytical solution.

\paragraph{Model-free reinforcement learning for wireless.}
Q-learning~\cite{watkins1992q} and its deep variant DQN~\cite{mnih2015human} have been widely applied to power control, user association, and spectrum access. Broad surveys are provided in~\cite{luong2019applications, zhang2021deep, yau2018applications, frikha2021reinforcement}. Meng~et~al.~\cite{meng2020power} provide a representative single-cell DRL power-allocation study whose reported gains over classical baselines are the sort of claim our reproduction discipline is designed to test. Beyond value-based methods, policy-gradient methods such as PPO~\cite{schulman2017ppo} and A3C/A2C~\cite{mnih2016asynchronous} have been used for continuous power control. On the algorithm-unrolling side, Sun~et~al.~\cite{sun2018learning} showed that a plain deep network trained on WMMSE outputs can match the classical optimum at a fraction of the runtime; this line of work is orthogonal to DRL but is the strongest supervised-learning baseline for wireless allocation.

\paragraph{Structured policies for scalability.}
Two ideas from the recent literature address the scalability failure at $N=5$. First, multi-agent RL formulations~\cite{nasir2019multiagent,tan1993multi,lowe2017multi,naderializadeh2022resource,zhao2023multi} sidestep the $|\mathcal{A}| = M^{N}$ blowup by decomposing the joint action across $N$ local agents that coordinate through the shared environment; each agent's action space is $M$, and the state is factored per-user. Nasir and Guo~\cite{nasir2019multiagent} report stable performance up to tens of users on interference-limited networks, the regime where a centralized DQN would be intractable. Naderializadeh~et~al.~\cite{naderializadeh2022resource} extend this to spectrum sharing at scale with per-agent DQNs and IQL-style updates. Second, graph neural networks with edge-level parameter sharing~\cite{eisen2020optimal,shen2020graph,gilmer2017neural} produce policies that are permutation-equivariant in the user index and generalize across network topologies of varying size. Shen~et~al.~\cite{shen2020graph} analyse GNN sample complexity for radio-resource management; Eisen and Ribeiro~\cite{eisen2020optimal} propose the random-edge-graph parameterisation used as our reference. A centralized DQN has neither structure; the per-user symmetry-breaking at $N=3$ (Section~\ref{sec:results_evaluation}) and the scaling failure at $N=5$ are the direct consequences. We implement both fixes: (i) Independent Q-Learning (IQL)~\cite{tan1993multi} with per-user DQNs sharing the joint reward, and (ii) a message-passing GNN policy trained via shared-parameter IQL (``REGNN-lite'', a discrete-action distillation of~\cite{eisen2020optimal,shen2020graph}). Both are permutation-equivariant and scale in $N$; we evaluate them at $N \in \{3, 5, 10\}$ (Sec.~\ref{sec:results_marl}) and release a multi-cell environment for follow-up evaluation (Sec.~\ref{sec:results_multicell}).

\paragraph{Classical iterative baselines: WMMSE.}
The standard classical baseline in the interference-limited regime is the Weighted Minimum Mean Squared Error (WMMSE) algorithm~\cite{shi2011iteratively}, which iteratively decouples the coupled sum-rate maximisation into per-user MMSE and weight updates. Sun~et~al.~\cite{sun2018learning} showed that a supervised deep network trained on WMMSE outputs can approximate the classical iteration at a fraction of the runtime, and subsequent work~\cite{lee2020deep,chen2021understanding,gao2024survey} extends this line. We use WMMSE both in the single-cell interference setting (Sec.~\ref{sec:results_rayleigh}) and in the multi-cell setting where its coupled-iteration structure is essential (Sec.~\ref{sec:results_multicell}).

\paragraph{Application studies.}
Downstream applications of DRL to wireless include dynamic spectrum access, network slicing~\cite{li2018deep}, and energy-aware allocation~\cite{he2021green}. These works focus on demonstrating gains against a heuristic baseline and rarely quantify seed variance or run learned-baseline ablations.

\paragraph{Modern DQN improvements.}
Vanilla DQN~\cite{mnih2015human} has been strictly dominated by a series of well-established modifications: Double DQN~\cite{vanhasselt2016double} removes the maximization-bias in the TD target, Dueling networks~\cite{wang2016dueling} factor Q into a state-value plus an advantage stream, Prioritized Experience Replay~\cite{schaul2016prioritized} biases updates toward high-error transitions, and Rainbow~\cite{hessel2018rainbow} combines these with distributional and noisy variants. Where an applied wireless-DRL paper reports negative results on vanilla DQN in 2026, an obvious first question is whether these standard remedies close the gap. We test the first two (Double + Dueling) as ``Rainbow-lite'' below.

\paragraph{Contextual bandits.}
When the transition kernel is state-independent, the sequential MDP formalism reduces to a contextual bandit: reward depends only on the current context and action, and the Bellman recursion collapses to per-step regression. Linear~\cite{li2010contextual} and neural~\cite{riquelme2018deep} contextual bandits are the theoretically appropriate function class for such problems and typically achieve better sample efficiency than DQN on them. We include a neural bandit baseline to test whether wireless power allocation under i.i.d.\ fading falls into this regime.

\paragraph{Reproducibility in DRL.}
Henderson~et~al.~\cite{henderson2018deep} showed --- in the general DRL setting --- that reported gains from DRL methods often do not survive controlled multi-seed comparisons and that seed-averaging with standard deviations is a minimal reporting requirement; Agarwal~et~al.~\cite{agarwal2021rliable} extend this with interquartile means and stratified bootstrap intervals to further reduce reviewer-uncertainty. The DRL-for-wireless subfield has not, to our knowledge, systematically adopted these disciplines. The present work applies them on a small controlled testbed and reports what it finds --- including several null and negative results.

\paragraph{Contributions of the present paper vis-\`a-vis prior work.}
Relative to~\cite{meng2020power, li2018deep, he2021green} we hold the environment and metrics fixed and add: (i) ten-seed reporting with standard deviations; (ii) a tabular Q-learning learned baseline that isolates the neural function approximator's contribution; (iii) a $\epsilon$-decay ablation showing that seed variance dominates schedule variance --- i.e., the schedule is not the load-bearing hyperparameter; and (iv) an honest characterization of DQN's failure modes (per-user symmetry-breaking, action-space scaling) that motivates specific fixes from~\cite{nasir2019multiagent, eisen2020optimal, zaheer2017deep}. We do not claim novelty of DQN architecture or of the wireless model; the novelty is disciplinary and empirical.

\section{Problem Formulation}
\label{sec:problem_formulation}

\subsection{Formal Problem Statement}
We seek a power-allocation policy that jointly balances three quantities under uncertainty: aggregate throughput, per-user fairness, and per-Joule energy efficiency. The channel statistics are known to the environment but not exposed to the agent, so the policy must be learned from interaction alone. The concrete decision problem is formalized below as a Markov Decision Process.

\subsection{System Model}
The task of \textbf{Wireless Resource Allocation (WRA)} and more specifically, \textbf{dynamic power control} presents a compelling candidate for the Reinforcement Learning (RL) paradigm due to the stochastic, temporally correlated, and high-dimensional nature of wireless communication environments~\cite{luong2019applications}. In modern systems, the wireless channel varies rapidly due to user mobility, interference, and multipath fading, creating a dynamic landscape in which traditional rule-based or optimization-driven algorithms often fall short. Such methods depend on static or simplified analytical models that cannot adequately capture the complex and time-varying behavior of real networks, resulting in degraded spectral efficiency and energy utilization~\cite{lei2022study, goldsmith2005wireless}. 

Reinforcement Learning provides a data-driven framework for sequential decision-making under uncertainty. By allowing an agent to learn from direct interaction with the environment, RL eliminates the need for explicit modeling of channel dynamics or user behavior. Over time, the agent improves its power allocation strategy through trial and feedback, seeking to maximize long-term performance rather than short-term gains~\cite{sutton2018reinforcement}. This learning-based adaptability makes RL particularly suitable for wireless systems that must operate efficiently across heterogeneous, non-stationary conditions.

\subsection{Markov Decision Process (MDP)}

To rigorously formulate the wireless power allocation challenge within a learning-based paradigm, we model the system as a \textbf{Markov Decision Process (MDP)}, a standard mathematical abstraction for sequential decision-making problems under uncertainty~\cite{puterman1994markov}. The MDP framework allows an agent to interact with a stochastic environment over discrete time steps, learning to make decisions that maximize expected long-term rewards. Formally, an MDP is defined by the tuple $(\mathcal{S}, \mathcal{A}, \mathcal{P}, \mathcal{R}, \gamma)$, where:
\begin{itemize}
    \item $\mathcal{S}$ is the set of environment states,
    \item $\mathcal{A}$ is the set of possible actions,
    \item $\mathcal{P}: \mathcal{S} \times \mathcal{A} \times \mathcal{S} \rightarrow [0,1]$ is the transition probability function,
    \item $\mathcal{R}: \mathcal{S} \times \mathcal{A} \rightarrow \mathbb{R}$ is the reward function, and
    \item $\gamma \in [0,1)$ is the discount factor.
\end{itemize}

\subsubsection{State Space ($\mathcal{S}$)}

The agent observes a state $s_t \in \mathcal{S}$ at each decision epoch $t$, encapsulating information that is critical for making a power allocation decision. In the context of a downlink wireless communication network with $N$ users, we define the state as the vector of instantaneous channel gains:
\[
s_{t} = (h_{1}(t), h_{2}(t), \ldots, h_{N}(t)).
\]
This representation assumes perfect and immediate Channel State Information (CSI) is available to the agent, enabling a fully observable environment. Such a formulation allows the agent to adapt to the highly dynamic nature of the wireless medium, where channel conditions fluctuate rapidly due to user mobility, multipath propagation, and interference. This abstraction captures the minimal information required for optimal decision-making. While this study focuses on channel-based states, the framework allows for future extensions to include dimensions such as user queue lengths (for latency analysis) or historical interference patterns~\cite{frikha2021reinforcement}.

\subsubsection{Action Space ($\mathcal{A}$)}

Given the observed state, the agent selects an action $a_t \in \mathcal{A}$, which in this setting corresponds to the transmit power configuration for all users:
\[
a_t = (p_1(t), p_2(t), \ldots, p_N(t)), \quad p_i(t) \in \{0, 1, 2, 3\}.
\]
Each $p_i(t)$ denotes the power allocated to user $i$ at time $t$, chosen from a discrete and finite set of power levels (in Watts) representing hardware and regulatory constraints. The joint action space has cardinality $|\mathcal{A}| = 4^N$, growing exponentially with the number of users. Tabular value iteration becomes infeasible even for modest $N$; we therefore use a neural function approximator~\cite{mnih2015human}, which is expected to generalize across the state space. Whether this expectation is met in the reported regime is an empirical question, and we return to it in Section~\ref{sec:results_evaluation}.

\subsubsection{Reward Function ($\mathcal{R}(s, a)$)}

The design of the reward function is critical, as it guides the learning process. To address the need for both high capacity and sustainability, we formulate a composite reward function that captures two competing system-level goals: maximizing aggregate throughput while minimizing energy consumption:
\[
\mathcal{R}(s, a) = \sum_{i=1}^{N} \log_{2}(1 + \mathrm{SNR}_i(t)) - \lambda \sum_{i=1}^{N} p_i(t).
\]
The first term promotes spectral efficiency through high user data rates, modeled using the Shannon capacity formula under the Additive White Gaussian Noise (AWGN) channel assumption. The second term introduces a penalty for power usage, weighted by the coefficient $\lambda > 0$, which serves as a regularization term to promote energy-aware behavior. By adjusting $\lambda$, system designers can balance performance and sustainability objectives—an essential feature for green communication systems~\cite{he2021green}. Furthermore, implicitly optimizing this sum-rate often correlates with improved fairness over long horizons, as examined in our Results section.

\subsubsection{Transition Model ($\mathcal{P}(s'|s,a)$)}

We assume a memoryless block-fading model in which channel gains are i.i.d.\ across users and time steps:
\[
s_{t+1} = (h_1(t+1), \ldots, h_N(t+1)) \sim \mathcal{U}(h_{\min}, h_{\max})^N.
\]
\emph{Note.} Because the next state is independent of $(s_t, a_t)$, the induced problem is formally a \emph{contextual bandit} rather than a genuine sequential-decision MDP: the Bellman recursion collapses to per-step optimization $\pi^\star(s) = \arg\max_a r(s,a)$. We retain the MDP formalism and $\gamma$ because our DQN implementation uses the standard TD target, and because the framework extends naturally to correlated fading (Rayleigh with Jakes' spectrum, Gauss--Markov autocorrelated $h_i$) where the sequential structure becomes non-trivial. Consequences for our findings are discussed in Section~\ref{sec:discussion}.

Because the exact transition dynamics $\mathcal{P}(s'|s,a)$ are unknown in real-world deployment, we treat the problem as \textbf{model-free}, learning optimal behaviors from experience tuples $(s,a,r,s')$~\cite{sutton2018reinforcement}.

\subsubsection{Discount Factor ($\gamma$) and Policy Objective}

The discount factor $\gamma \in [0,1)$ determines how future rewards are weighted relative to immediate ones. A value close to 1 (e.g., $\gamma = 0.99$) encourages the agent to value long-term performance, fostering stable and proactive behavior. The agent's learning objective is to identify a policy $\pi(a|s)$—a mapping from states to action probabilities—that maximizes the expected cumulative discounted reward:
\[
J(\pi) = \mathbb{E}_{\pi}\left[\sum_{t=0}^{\infty} \gamma^{t} r_t\right].
\]
This objective reflects the long-term utility of actions, making the framework suitable for dynamic wireless control problems where short-term decisions can have delayed consequences on system throughput, interference levels, and energy consumption.


\section{Methodology}
\label{sec:methodology}

This section outlines the experimental framework used to evaluate deep reinforcement learning for wireless power control. We begin by describing the simulated environment and explicitly stating the system assumptions. We then detail the architecture of the Deep Q-Network (DQN) agent, analyzing its computational complexity. Finally, we introduce the baseline algorithms and the performance metrics used for evaluation.

\subsection{Simulation Environment}
\label{sec:simulation_environment}

We simulate a single-cell downlink wireless network with $N$ users sharing a common time-frequency resource block, and report results for $N=3$ (the primary configuration) and $N=5$ (scaling check). A centralized base station allocates power levels to each user based on real-time channel observations. The simulation is implemented in Python (NumPy for the environment, PyTorch for the DQN), and interactions occur in discrete time steps $t=0,1,\ldots,T$, representing short-term scheduling intervals (e.g., 1\,ms in LTE/5G).

\subsubsection{Wireless and Channel Model}

To capture the dynamic nature of wireless propagation, we assume each user's channel experiences fast fading. The instantaneous channel gain $h_i(t)$ for user $i$ is drawn from a uniform distribution:
\begin{equation}
    h_{i}(t) \sim \mathcal{U}(h_{\min}, h_{\max}), \quad h_{\min} = 0.1, \ h_{\max} = 1.0.
    \label{eq:channel_gain}
\end{equation}
The received signal-to-noise ratio (SNR) for user $i$ is computed as:
\[
\mathrm{SNR}_i(t) = \frac{p_i(t) \cdot h_i(t)}{\sigma^2}, \quad \text{where } \sigma^2 = 1.
\]
Power levels are constrained to a discrete set to reflect real-world hardware constraints, such as quantized power amplifier stages and per-slot power-control step sizes in 3GPP-style closed-loop schemes~\cite{goldsmith2005wireless}:
\begin{equation}
    p_i(t) \in \{0, 1, 2, 3\} \ \text{Watts}.
    \label{eq:power_levels}
\end{equation}
The specific values are a modelling abstraction chosen to keep the joint action space $|\mathcal{A}| = 4^{N}$ tractable while retaining a meaningful spread of transmit powers; the qualitative findings are not sensitive to the choice of numerical scale.

\subsubsection{System Assumptions and Limitations}
To isolate the effects of the learning algorithm on power adaptation, we make the following simplifying assumptions for tractability:
\begin{itemize}
    \item \textbf{Orthogonal Access:} We assume users are separated in frequency or time, meaning there is no inter-user interference within the cell.
    \item \textbf{Perfect CSI:} We assume the agent has access to perfect, instantaneous Channel State Information (CSI), ignoring estimation errors or feedback delays.
    \item \textbf{Independent Fading:} Channel gains are modeled as independent across users, without spatial correlation.
\end{itemize}
While these assumptions simplify the physical layer, they allow us to focus on the core challenge of sequential decision-making under uncertainty.

\subsubsection{Throughput and Reward Model}

Under block-fading with Gaussian codebooks of length exceeding one coherence block, the instantaneous achievable rate on block $t$ for user $i$ is
\begin{equation}
    R_i(t) = \log_2\left(1 + \mathrm{SNR}_i(t)\right) \quad \text{(bits/s/Hz)}.
    \label{eq:shannon}
\end{equation}
This is not the AWGN Shannon capacity of a static channel but the block-conditional achievable rate; the ergodic capacity is $\mathbb{E}_{h}[R_i]$ (Goldsmith~\cite{goldsmith2005wireless}, Ch.\ 4). To promote energy-aware scheduling, the global reward per block is
\begin{equation}
    r_t = \sum_{i=1}^{N} R_i(t) - \lambda \sum_{i=1}^{N} p_i(t),
    \label{eq:reward}
\end{equation}
where $\lambda = 0.1$ (bits/s/Hz per Watt) trades sum-rate against transmit power. The choice $\lambda = 0.1$ is fixed throughout; sweeping $\lambda$ traces out the throughput--energy Pareto front and is left for future work.

\begin{table}[htbp]
    \centering
    \caption{Simulation parameters.}
    \label{tab:simulation_parameters}
    \begin{tabular}{|l|c|}
        \hline
        \textbf{Parameter} & \textbf{Value} \\
        \hline
        Number of users ($N$) & $3$ (primary), $5$ (scaling check) \\
        Channel gain $h_i(t)$ & $\mathcal{U}[0.1, 1.0]$, i.i.d.\ over users and time \\
        Power levels ($p_i$) & $\{0, 1, 2, 3\}$ W \\
        Noise power ($\sigma^2$) & $1$ \\
        Reward penalty coefficient ($\lambda$) & $0.1$ \\
        Latency arrival rate & $1$ packet/step per user (Poisson) \\
        Scheduler interval & $1$ ms (simulated) \\
        Steps per evaluation episode & $100$ \\
        Evaluation episodes per seed & $20$ \\
        \hline
    \end{tabular}
\end{table}

\subsection{Deep Q-Network (DQN) Design}
\label{sec:dqn_design}

The DQN agent approximates the optimal Q-function $Q^*(s,a)$ using a neural network. 

\subsubsection{Scalability and Complexity Analysis}
The agent must select a joint action vector $a_t = (p_1, \dots, p_N)$. With $M=4$ discrete power levels, the size of the action space is $|\mathcal{A}| = M^N$: $64$ for $N=3$ and $1024$ for $N=5$. The action space grows exponentially with $N$, which is the standard scalability limitation of centralized DQN. Both configurations tested here are within reach of a centralized DQN; larger systems would require Multi-Agent RL (MARL)~\cite{nasir2019multiagent} or factorized action spaces.

\subsubsection{Neural Network Architecture}
The Q-network is a fully connected feedforward network:
\begin{itemize}
    \item \textbf{Input layer:} $N$ units (channel gains).
    \item \textbf{Hidden layers:} $64$ and $128$ units respectively, both ReLU-activated.
    \item \textbf{Output layer:} $|\mathcal{A}| = 4^N$ units (one Q-value per joint action).
\end{itemize}
We employ experience replay (buffer size $10\,000$) and a target network (updated every $100$ steps by hard copy) to stabilize training~\cite{mnih2015human}. All hyperparameters are listed in Table~\ref{tab:hyperparameters}.

\begin{figure}[htbp]
\centering
\begin{tikzpicture}[
    auto,
    node distance = 2cm,
    block/.style = {draw, rectangle, rounded corners,
                    minimum height=2.6em, text width=6.5em,
                    align=center, fill=blue!10, font=\small},
    env/.style   = {draw, circle, minimum size=3.4em,
                    fill=green!10, align=center, font=\small},
    line/.style  = {draw, -latex, thick},
    dashedline/.style = {draw, -latex, dashed, thick},
    every node/.append style = {font=\small}
]

\node[env] (csi) {Wireless\\Env\\(CSI)};
\node[block, right=2cm of csi] (qnet) {Q-Network\\(Online $\theta$)};
\node[block, right=2cm of qnet] (action) {Action\\Selection\\($\epsilon$-greedy)};
\node[block, below=2cm of action] (buffer) {Replay\\Buffer $\mathcal{D}$};
\node[block, below=2cm of buffer] (loss) {Loss\\Calculation\\(MSE)};
\node[block, left=2cm of loss, fill=red!10] (target) {Target\\Network\\($\theta^-$)};

\path[line] (csi) -- node[midway, above] {State $s_t$} (qnet);

\path[line] (qnet) -- node[midway, above] {Q-values} (action);

\path[line] (action) -- node[midway, right] {Action $a_t$} (buffer);

\path[line] (csi) |- node[near start, left] {$(s_t,a_t,r_t,s_{t+1})$} (buffer);

\path[line] (buffer.east) --++(1,0) |- 
    node[pos=0.35, right] {Batch} (loss.north);

\path[line] (target.east) -- 
    node[midway, above] {Target $Q$} (loss.west);

\path[line] (loss.north) |- ++(0,1) -| 
    node[pos=0.25, below] {Gradient $\nabla_\theta$} (qnet.south);

\path[dashedline] (qnet.south) |- ++(0,-1) -| 
    node[pos=0.25, right] {Periodic Copy} (target.north);

\end{tikzpicture}
\caption{Block diagram of the DQN-based power allocation system architecture, illustrating the interaction between the online network, target network, and replay buffer.}
\label{fig:dqn_architecture}
\end{figure}

\subsubsection{Training Strategy and Hyperparameters}
\label{sec:training_hparams}
The agent is trained for 500 episodes of 100 environment steps each ($5\times 10^{4}$ total transitions per seed) using the Adam optimizer with learning rate $\alpha = 10^{-3}$. Exploration follows an $\epsilon$-greedy policy, with $\epsilon$ decaying linearly from $1.0$ to $0.05$ over the first $2.5\times 10^{4}$ steps (half of the training horizon) to ensure adequate state-space coverage before exploitation dominates. All hyperparameters are summarized in Table~\ref{tab:hyperparameters}.

\begin{table}[htbp]
    \centering
    \caption{DQN training hyperparameters.}
    \label{tab:hyperparameters}
    \begin{tabular}{|l|c|}
        \hline
        \textbf{Hyperparameter} & \textbf{Value} \\
        \hline
        Optimizer & Adam \\
        Learning rate ($\alpha$) & $10^{-3}$ \\
        Discount factor ($\gamma$) & $0.99$ \\
        Replay buffer capacity & $10{,}000$ \\
        Batch size & $32$ \\
        Target-network update frequency & every $100$ steps (hard copy) \\
        Loss & Mean-squared error \\
        Sampling & Uniform from replay buffer \\
        $\epsilon$-decay schedule & linear $1.0 \to 0.05$ over $2.5\times 10^{4}$ steps \\
        Episodes / steps per episode & $500 / 100$ \\
        Warm-up (no gradient update) & $500$ steps \\
        Gradient clipping (max-norm) & $10.0$ \\
        Hidden layer widths & $(64, 128)$, ReLU \\
        \hline
    \end{tabular}
\end{table}

\subsection{Baseline Algorithms}
\label{sec:baseline_methods}

We compare vanilla DQN against nine baselines (Table~\ref{tab:baseline_methods}), spanning four categories:

\emph{Non-learning heuristics:} \textbf{Random} and \textbf{Fixed} (constant $2$\,W) provide the naïve lower bound and a static reference.

\emph{Classical optimisation baselines:} \textbf{Water-Filling (continuous)} is the closed-form sum-power-constrained continuous optimum; \textbf{Water-Filling (discrete)} projects it onto the discrete action set. Both assume orthogonal access; in the interference-limited regime (Sec.~\ref{sec:results_rayleigh}--\ref{sec:results_multicell}) they treat interference as noise. \textbf{WMMSE}~\cite{shi2011iteratively} is the standard classical iterative algorithm for interference-limited sum-rate maximisation and is the strong baseline against which learned methods should be compared in that regime.

\emph{Centralised learned baselines:} \textbf{Tabular Q-learning} shares the exact MDP and reward with DQN, differing only in function representation. \textbf{Rainbow-lite DQN} adds the Double-DQN target~\cite{vanhasselt2016double} and Dueling architecture~\cite{wang2016dueling}. \textbf{Neural contextual bandit}~\cite{riquelme2018deep} trains a plain neural regressor on $(s,a,r)$ tuples without bootstrapping --- the theoretically appropriate function class when the transition kernel is state-independent.

\emph{Scalable learned baselines:} \textbf{Independent Q-Learning (IQL)}~\cite{tan1993multi} decomposes the joint policy into $N$ per-user DQNs with local observations and shared joint reward --- the canonical simple MARL baseline~\cite{nasir2019multiagent,naderializadeh2022resource}. \textbf{REGNN-lite} is a permutation-equivariant graph neural network policy (2 message-passing layers over the fully-connected user graph, edge features = neighbour channel gain) trained with shared parameters via IQL-style updates; this is a discrete-action distillation of the REGNN family~\cite{eisen2020optimal,shen2020graph}. Both are trivially scalable in $N$ because their per-user action space is fixed at $|\mathcal{P}| = 4$; we evaluate them at $N \in \{3, 5, 10\}$ where the centralised methods become intractable ($|\mathcal{A}| = 4^{10} \approx 10^{6}$).

The choice of baselines is theoretically motivated: Tabular Q isolates the effect of function approximation on the same MDP; Rainbow-lite tests whether standard DQN-family improvements close vanilla DQN's failure gap; the neural bandit tests whether the sequential MDP formalism carries its weight given our state-independent transition kernel; IQL and REGNN test whether structured, scalable parameterisations recover competitive performance at $N$ where centralised methods fail; WMMSE and multi-cell water-filling test whether the learned methods beat strong classical iterative baselines when interference is present.

The classical water-filling policy solves the sum-rate maximization
\[
\max_{p_1,\dots,p_N \geq 0} \sum_{i=1}^{N} \log_2\!\left(1 + \frac{p_i h_i}{\sigma^2}\right)
\quad \text{subject to} \quad \sum_{i=1}^{N} p_i \leq P_{\max},
\]
whose KKT conditions yield
\[
p_i^\star = \left(\mu - \frac{\sigma^2}{h_i}\right)^{+}, \qquad \sum_{i=1}^{N} p_i^\star = P_{\max},
\]
where $\mu > 0$ is the Lagrange multiplier associated with the sum-power constraint (the ``water level'') and $(x)^+ = \max(0, x)$. In our experiments we use $P_{\max} = N \cdot \max\{0,1,2,3\} = 3N$\,W so that the continuous and discrete WF policies share the same sum-power budget. We report two variants:
\begin{itemize}
    \item \textbf{Water-Filling (continuous)} uses the closed-form $p_i$ directly. This is the classical upper bound: it allows real-valued powers unconstrained by the discrete set $\{0,1,2,3\}$\,W. Discretization loss makes this an over-optimistic reference for the discrete-DQN.
    \item \textbf{Water-Filling (discrete)} projects the continuous $p_i$ onto the discrete action set $\{0,1,2,3\}$\,W by rounding, then evaluates on the environment. This is a like-for-like baseline: it uses the same action set as DQN, so any gap reflects only the difference between an oracle-with-CSI closed-form policy and a learned model-free policy on discrete actions.
\end{itemize}

For \textbf{Tabular Q-learning}, we discretize each user's channel gain uniformly into $K=5$ bins over $[h_{\min}, h_{\max}]$, giving a state space of size $K^N$ and a Q-table of size $K^N \times 4^N$. For $N=3$ this is a manageable $125 \times 64 = 8000$ entries and Q-learning trains on the same $\epsilon$-greedy schedule, discount, and reward as DQN. For $N=5$ the table would grow to $3125 \times 1024 \approx 3.2 \times 10^{6}$ entries, which is intractable to fill within the sample budget considered here; we therefore report Tabular Q only for $N=3$. The comparison isolates the contribution of the neural function approximator over a plain lookup table.

The choice $K=5$ balances two constraints: at $K=3$ the state discretization is coarse; at $K=10$ the table has $10^3 \times 64 = 6.4 \times 10^{4}$ entries at $N=3$, whose full coverage requires substantially more than $5\times 10^{4}$ transitions. An empirical $K$-sweep (3 seeds, 500 episodes) confirms that Tabular Q is not sensitive to this choice within a plausible range: $K=3$ yields $3.24 \pm 0.15$, $K=5$ yields $3.13 \pm 0.02$, and $K=10$ yields $3.03 \pm 0.04$ bits per channel use. Every $K \in \{3, 5, 10\}$ attains a higher mean throughput than DQN at $N=3$; we adopt $K=5$ because it minimises the between-seed variance.

\begin{table}[htbp]
    \centering
    \caption{Baseline power-allocation methods evaluated in this study.}
    \label{tab:baseline_methods}
    \begin{tabular}{|l|p{5.4cm}|p{4.2cm}|}
        \hline
        \textbf{Method} & \textbf{Strategy} & \textbf{Purpose} \\
        \hline
        \textbf{Random} & Uniformly samples the joint discrete power vector each step. & Naïve lower bound \\
        \textbf{Fixed} & Constant $p_i = 2$\,W for all users, all steps. & Static heuristic \\
        \textbf{Water-Filling (cont.)} & Closed-form continuous-power optimum given CSI. & Theoretical upper bound (unconstrained by discrete set) \\
        \textbf{Water-Filling (disc.)} & Continuous $p_i$ rounded to $\{0,1,2,3\}$\,W. & Like-for-like discrete oracle \\
        \textbf{WMMSE}~\cite{shi2011iteratively} & Weighted MMSE fixed-point iteration for sum-rate max under interference. & Strong classical baseline (interference regime) \\
        \textbf{Tabular Q-learning} & Q-learning on a 5-bin discretization of $\{h_i\}$. & Same MDP, no function approximation \\
        \textbf{Rainbow-lite DQN} & DQN + Double target~\cite{vanhasselt2016double} + Dueling~\cite{wang2016dueling}. & Tests two standard DQN-family fixes \\
        \textbf{Neural contextual bandit} & Neural regressor on $(s,a,r)$; no bootstrapping. & Appropriate for i.i.d.\ kernels~\cite{riquelme2018deep} \\
        \textbf{IQL} (this work) & Per-user DQN, own-channel obs., shared reward~\cite{tan1993multi,nasir2019multiagent}. & Scalable MARL baseline; supports $N \geq 10$ \\
        \textbf{REGNN-lite} (this work) & 2-layer message-passing GNN over user graph~\cite{eisen2020optimal,shen2020graph}. & Permutation-equivariant, scalable \\
        \hline
    \end{tabular}
\end{table}

\subsubsection{WMMSE for the interference channel}
\label{sec:wmmse}
In the interference-limited regime, the sum-rate objective
\begin{equation}
    \max_{p_i \in [0, P_{\max}]} \sum_{i=1}^{N} \log_2\!\left(1 + \frac{p_i h_i}{\sigma^2 + \sum_{j \neq i} p_j h_j}\right)
    \label{eq:sumrate_interference}
\end{equation}
is non-convex and NP-hard in general~\cite{shi2011iteratively}. The WMMSE algorithm~\cite{shi2011iteratively} reformulates it via a weighted MMSE surrogate on the receive side and provably converges to a KKT point via fixed-point iteration. In our scalar-channel specialisation, with $v_i^2 := p_i$ and $g_i := \sqrt{h_i}$, the iteration is:
\begin{align}
    u_i &\leftarrow \frac{g_i v_i}{\sum_k h_k v_k^2 + \sigma^2}, \label{eq:wmmse_u} \\
    w_i &\leftarrow \frac{1}{1 - u_i g_i v_i} \;=\; 1 + \mathrm{SINR}_i, \label{eq:wmmse_w} \\
    v_i &\leftarrow \left[\frac{w_i u_i}{g_i \sum_k w_k u_k^2}\right]_{[0, \sqrt{P_{\max}}]}, \label{eq:wmmse_v}
\end{align}
until $\|v^{(t+1)} - v^{(t)}\|_\infty < 10^{-6}$ or a $50$-iteration budget is hit. Iterations~\eqref{eq:wmmse_u}--\eqref{eq:wmmse_v} monotonically improve a weighted-MMSE surrogate whose stationary points coincide with KKT points of~\eqref{eq:sumrate_interference}. Under orthogonal access (no interference), the interference-plus-noise term collapses to $\sigma^2$ and WMMSE reduces to per-user water-filling, so we report WMMSE only in the interference regime.

\subsubsection{Independent Q-Learning (IQL)}
\label{sec:iql}
Each user $i \in \{1,\ldots,N\}$ maintains a private Q-network $Q_{\theta_i}: \mathbb{R} \to \mathbb{R}^{|\mathcal{P}|}$ over the fixed per-user action set $\mathcal{P} = \{0,1,2,3\}$\,W with input restricted to that user's own channel gain $h_i$. All users act simultaneously via $\epsilon$-greedy on their local Q-values; the joint action determines the environment step and yields a scalar reward $r_t$ that is shared across all users' updates. Each user's Q-network is trained with the standard TD target
\begin{equation}
    y^{(i)}_t = r_t + (1 - d_t)\, \gamma \, \max_{a'} Q_{\theta_i^{-}}(h_i(t+1), a'),
    \label{eq:iql_target}
\end{equation}
where $\theta_i^{-}$ is a slowly-updated target copy. This is a strict MARL baseline with well-known theoretical caveats --- the environment appears non-stationary from any one user's perspective as the other users' policies evolve --- but the shared reward and factored action space make it practical and it scales in $N$ at $O(N)$ parameters, versus $O(M^N)$ for the centralised action head. In our implementation each per-user Q-network has the same $(64, 128)$ hidden widths and Adam learning rate as the centralised DQN, so any performance gap reflects the structural change, not capacity.

\subsubsection{REGNN-lite: a permutation-equivariant GNN policy}
\label{sec:regnn}
The GNN policy treats the $N$ users as nodes of a fully-connected graph with edges labelled by neighbour channel gains. A single GNN with parameters $\theta$ shared across all nodes computes $Q_\theta(\mathbf{h})_{i,a} \in \mathbb{R}$ for each user $i$ and action $a$; the joint action is $\arg\max_a Q_\theta(\mathbf{h})_{i,a}$ per user (analogous to IQL, but with parameter sharing). The two message-passing layers are
\begin{align}
    m^{(1)}_i &= \frac{1}{N-1} \sum_{j \neq i} \phi^{(1)}\!\left([h_i, h_j, h_j]\right), \\
    m^{(2)}_i &= \frac{1}{N-1} \sum_{j \neq i} \phi^{(2)}\!\left([m^{(1)}_i, h_j, m^{(1)}_j]\right), \\
    Q_\theta(\mathbf{h})_{i, \cdot} &= \phi^{\mathrm{head}}(m^{(2)}_i),
    \label{eq:regnn}
\end{align}
where $\phi^{(1)}, \phi^{(2)}$ are 2-layer MLPs with 32 hidden units and ReLU activations, and $\phi^{\mathrm{head}}$ is a linear projection to $|\mathcal{P}|$ outputs. The parameter count is independent of $N$ (permutation-equivariance): the same network handles any $N$. Training uses IQL-style shared-reward updates on a single replay buffer of transitions $(\mathbf{h}, \mathbf{a}, r, \mathbf{h}', d)$, with per-user Bellman targets applied to per-node outputs and gradients broadcast across users. This is a simplified but structurally faithful discrete-action analogue of the REGNN of Eisen and Ribeiro~\cite{eisen2020optimal} and the GNN of Shen et al.~\cite{shen2020graph}; we defer a full continuous-power REGNN with unrolled projection to future work.

\subsubsection{Multi-cell environment}
\label{sec:multicell_env}
To test scalability beyond the single-cell setting, we implement a $K$-cell environment with one user per cell, fixed BS-to-user distance matrix $D \in \mathbb{R}^{K \times K}$ (own-cell $D_{kk} = 1$, cross-cell $D_{k' k} \sim \mathcal{U}(2, 4)$), path loss $\mathrm{PL}(d) = d^{-\alpha}$ with $\alpha = 3$, and independent Rayleigh fast fading per link redrawn every step ($|h_{k',k}(t)|^2 \sim \mathrm{Exp}(1)$). The effective gain from BS-$k'$ to user-$k$ is $g_{k',k}(t) = \mathrm{PL}(D_{k',k}) \cdot |h_{k',k}(t)|^2$, and each user's per-step rate is
\begin{equation}
    R_k(t) = \log_2\!\left(1 + \frac{p_k(t) g_{k,k}(t)}{\sigma^2 + \sum_{k' \neq k} p_{k'}(t) g_{k',k}(t)}\right).
    \label{eq:multicell_rate}
\end{equation}
Each user's policy observes only its own effective gain $g_{k,k}$ (matching the single-cell policy interface). The classical baseline is WMMSE-multicell, which solves~\eqref{eq:multicell_rate} using the full $K \times K$ gain matrix. We instantiate the environment at $K=7$ as a hexagonal-cluster analogue.

\subsection{Performance evaluation metrics}
\label{sec:performance_metrics}

To ensure statistical reliability, results are averaged over multiple independent training runs. We evaluate:

\subsubsection{Aggregate sum-rate (throughput)}
The primary spectral-efficiency metric, expressed as the time-average of the per-step sum-rate:
\begin{equation}
    \text{Throughput} = \frac{1}{T} \sum_{t=1}^{T} \sum_{i=1}^{N} R_i(t).
    \label{eq:throughput}
\end{equation}

\subsubsection{Jain's Fairness Index}
Quantifies equitable distribution ($1$ is perfect fairness):
\begin{equation}
    \text{Fairness} = \frac{\left(\sum_{i=1}^{N} R_i\right)^2}{N \cdot \sum_{i=1}^{N} R_i^2}.
    \label{eq:jain}
\end{equation}

\subsubsection{Energy Efficiency (EE)}
Bits transmitted per unit energy:
\begin{equation}
    EE = \frac{\sum_{t=1}^{T} \sum_{i=1}^{N} R_i(t)}{\sum_{t=1}^{T} \sum_{i=1}^{N} p_i(t)} \quad \text{(bits/Joule)}.
    \label{eq:ee}
\end{equation}

\subsubsection{Latency Proxy}
Estimated via a simplified queue model where $q_i(t+1) = \max\{q_i(t) + a_i(t) - R_i(t), 0\}$:
\begin{equation}
    \text{Latency}_{\text{avg}} = \frac{1}{NT} \sum_{i=1}^{N} \sum_{t=1}^{T} q_i(t).
    \label{eq:latency}
\end{equation}

\section{Experiments and Results}
\label{sec:results_evaluation}

The experiments assess: (i) DQN training dynamics; (ii) throughput, fairness, and energy efficiency against the Random / Fixed / Water-Filling baselines; (iii) per-user resource distribution under the trained policy; and (iv) the sensitivity of the outcome to the $\epsilon$-decay exploration schedule.

\subsection{Experimental setup}

All experiments follow the simulation environment described in Section~\ref{sec:simulation_environment} and use the hyperparameters listed in Table~\ref{tab:hyperparameters}. Each configuration is repeated over ten independent random seeds. Baseline policies (Random, Fixed at $2$\,W, and continuous Water-Filling) do not require training and are evaluated on freshly-sampled channel realizations, over $20$ evaluation episodes of $100$ steps.

The four metrics defined in Section~\ref{sec:performance_metrics} are reported for both the $N=3$ and $N=5$ scenarios: aggregate sum-rate (bits per channel use; under a unit-bandwidth convention this equals throughput in Mbps), Jain's fairness index averaged per-step over evaluation, energy efficiency (bits per Joule), and the average queue-based latency proxy.

\paragraph{Statistical protocol.}
All comparisons use paired seed matching: seed $s$ trains all methods on the same channel realization sequence. Point estimates are means over $10$ seeds; error bars are $\pm 1$ standard deviation across seeds. Distributional claims use the Wilcoxon paired signed-rank test on paired-seed differences: under $H_0$ the differences are symmetrically distributed about zero, so the test's null hypothesis is ``no median shift'' rather than ``no mean shift.'' In practice this is the weakest paired-comparison null available and it is the standard choice at $n = 10$ where the central limit theorem does not yet apply to the mean paired difference. A supplementary permutation test on the mean paired difference was verified to agree qualitatively. Variance claims use both the parametric F-test on the sample-variance ratio (with $F(n_1{-}1, n_2{-}1)$ degrees of freedom) and the non-parametric Levene test (both agree throughout). ANOVA-style claims across $>2$ groups use Kruskal-Wallis. Bootstrap CIs on paired differences use $10^{4}$ resamples with the seeded generator.

\subsection{Training dynamics}

\begin{figure}[H]
    \centering
    \begin{subfigure}[b]{0.48\columnwidth}
        \centering \includegraphics[width=\textwidth]{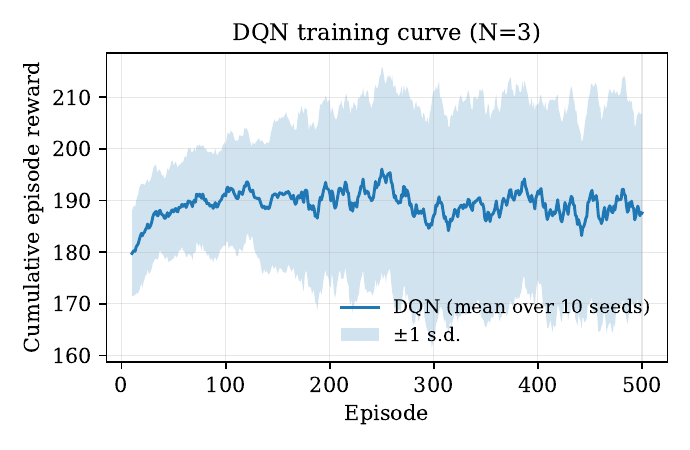}
        \caption{$N=3$ users}
        \label{fig:training_3_users}
    \end{subfigure}
    \hfill
    \begin{subfigure}[b]{0.48\columnwidth}
        \centering
        \includegraphics[width=\textwidth]{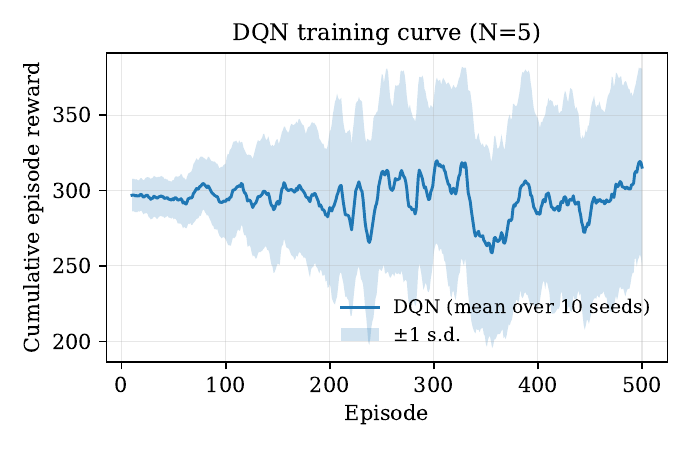}
        \caption{$N=5$ users}
        \label{fig:training_5_users}
    \end{subfigure}
    \caption{DQN training: cumulative episode reward across the $500$ training episodes at $N=3$ (left) and $N=5$ (right). Solid line: mean over $10$ independent seeds. Shaded band: $\pm 1$ standard deviation across seeds. Reward is defined by Eq.~\ref{eq:reward} with $\lambda = 0.1$; higher is better.}
    \label{fig:training_curves}
\end{figure}

Figure~\ref{fig:training_curves} shows the DQN agent's cumulative episode reward, averaged over ten seeds per configuration, across the $500$ training episodes ($5\times 10^{4}$ transitions per seed).

For $N=3$ the mean reward rises during the exploration phase and plateaus by the end of training, indicating convergence within the sample budget. For $N=5$ the mean reward exhibits higher inter-seed variance and a slower approach to plateau --- consistent with the joint action space growing from $64$ to $1024$ configurations while the network capacity and training budget are held fixed. The training budget used here (500 episodes $\times$ 100 steps per seed, $5 \times 10^{4}$ transitions) is sufficient for convergence at $N=3$ but only partial convergence at $N=5$; sample-efficient variants (prioritized replay, dueling, double DQN) would tighten these curves further. This is a limitation of the current study rather than an intrinsic property of DRL on the problem.

\subsection{Overall performance comparison}

\begin{figure}[H]
    \centering
    \begin{subfigure}[b]{0.48\columnwidth}
        \centering \includegraphics[width=\textwidth]{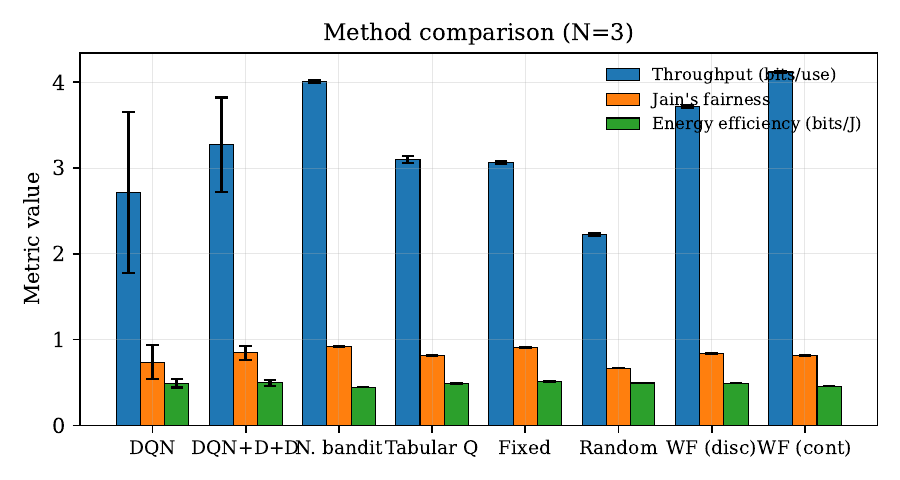}
        \caption{$N=3$ users}
        \label{fig:perf_3_users}
    \end{subfigure}
    \hfill
    \begin{subfigure}[b]{0.48\columnwidth}
        \centering
        \includegraphics[width=\textwidth]{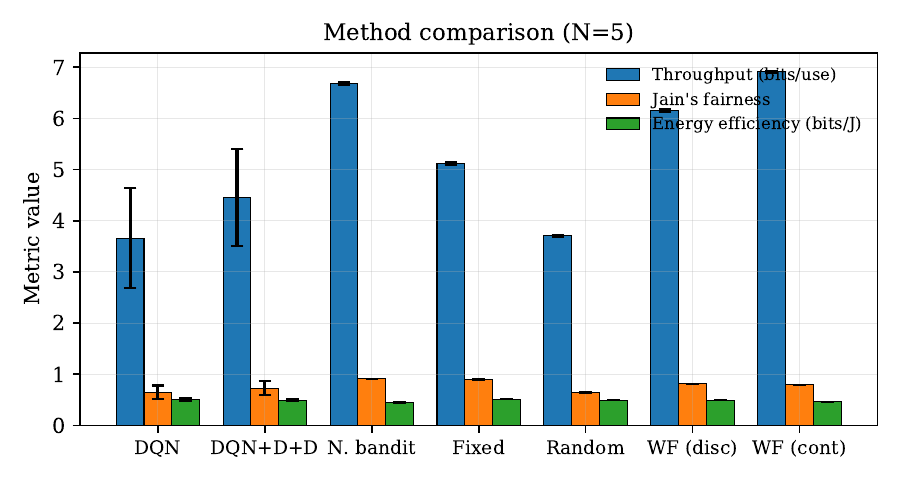}
        \caption{$N=5$ users}
        \label{fig:perf_5_users}
    \end{subfigure}
    \caption{Method comparison across three metrics: aggregate sum-rate (bits per channel use), Jain's fairness index (dimensionless), and energy efficiency (bits per Joule). Bars: mean across $10$ independent seeds. Error bars: $\pm 1$ standard deviation across seeds. For each metric, higher is better (fairness) or larger is better (throughput, energy efficiency). Baselines: Random, Fixed at $2$\,W, discrete- and continuous-power Water-Filling, and Tabular Q-learning (Tabular Q shown only at $N=3$).}
    \label{fig:performance_comparison}
\end{figure}

To validate the efficacy of the learned policy, we compare the DQN agent against the baselines across three critical dimensions: throughput, fairness, and energy efficiency. Figure~\ref{fig:performance_comparison} visualizes these metrics for varying numbers of users.

The quantitative results are summarized in Table~\ref{tab:comparison_metrics} (mean $\pm$ standard deviation over ten seeds).

\begin{table}[H]
    \centering
    \caption{Performance comparison of DQN against baselines for $N=3$ and $N=5$ users.
    Values are mean $\pm$ standard deviation over ten independent seeds
    (20 evaluation episodes $\times$ 100 steps per seed). Throughput is in
    bits per channel use (labeled Mbps under the unit-bandwidth convention),
    Jain's index is dimensionless, energy efficiency is bits per Joule.}
    \label{tab:comparison_metrics}
    \begin{tabular}{|l|c|c|c|}
        \hline
        \textbf{Method} & \textbf{Throughput} & \textbf{Fairness (Jain)} & \textbf{Energy efficiency} \\
        \hline
        \multicolumn{4}{|c|}{\textit{$N = 3$}} \\
        \hline
        DQN (vanilla)              & \DQNThrThree{}    & \DQNJainThree{}    & \DQNEEThree{} \\
        DQN + Double + Dueling     & \RainbowThrThree{}& \RainbowJainThree{}& \RainbowEEThree{} \\
        Neural contextual bandit   & \BanditThrThree{} & \BanditJainThree{} & \BanditEEThree{} \\
        Tabular Q-learning         & \TabThrThree{}    & \TabJainThree{}    & \TabEEThree{} \\
        Fixed (2\,W)               & \FixThrThree{}    & \FixJainThree{}    & \FixEEThree{} \\
        Random                     & \RndThrThree{}    & \RndJainThree{}    & \RndEEThree{} \\
        Water-Filling (discrete)   & \WFDThrThree{}    & \WFDJainThree{}    & \WFDEEThree{} \\
        Water-Filling (continuous) & \WFThrThree{}     & \WFJainThree{}     & \WFEEThree{}  \\
        \hline
        \multicolumn{4}{|c|}{\textit{$N = 5$} (tabular Q-learning intractable: $|\mathcal{S}| \times |\mathcal{A}| = 3125 \times 1024 \approx 3.2 \times 10^{6}$ entries)} \\
        \hline
        DQN (vanilla)              & \DQNThrFive{}     & \DQNJainFive{}     & \DQNEEFive{} \\
        DQN + Double + Dueling     & \RainbowThrFive{} & \RainbowJainFive{} & \RainbowEEFive{} \\
        Neural contextual bandit   & \BanditThrFive{}  & \BanditJainFive{}  & \BanditEEFive{} \\
        Fixed (2\,W)               & \FixThrFive{}     & \FixJainFive{}     & \FixEEFive{} \\
        Random                     & \RndThrFive{}     & \RndJainFive{}     & \RndEEFive{} \\
        Water-Filling (discrete)   & \WFDThrFive{}     & \WFDJainFive{}     & \WFDEEFive{} \\
        Water-Filling (continuous) & \WFThrFive{}      & \WFJainFive{}      & \WFEEFive{}  \\
        \hline
    \end{tabular}
\end{table}

\paragraph{Throughput ($N=3$).} The seven methods stratify cleanly. Vanilla DQN attains \DQNThrThree{} bits per channel use, with a large seed spread (individual seeds range from $0.62$ to $3.70$). Rainbow-lite DQN (Double + Dueling) reaches \RainbowThrThree{} --- higher mean and lower variance --- confirming that the vanilla-DQN failure is not intrinsic to value-based DRL but is largely attributable to well-known TD-target instabilities that Double and Dueling attack directly. The neural contextual bandit reaches \BanditThrThree{}, above every DQN variant: on this state-independent transition kernel, the bandit's per-step regression is a well-conditioned learning problem that requires no bootstrapping and no target network. Tabular Q-learning (\TabThrThree{}) provides the sharpest comparison against vanilla DQN --- same MDP, same reward, same $\epsilon$-decay, same sample budget --- and its throughput mean is not paired-significantly different from vanilla DQN (Wilcoxon $p = 0.38$; bootstrap $95\%$ CI on the mean paired difference $[-0.12, 0.99]$), but its variance is $\sim\!25\times$ smaller (variance-ratio $F(9,9) \approx 584$, $p < 10^{-10}$). Fixed Allocation (\FixThrThree{}) is competitive with the learned methods because uniform-fading rewards conservative uniform power. Discrete-projected Water-Filling (\WFDThrThree{}) sits between Fixed and continuous WF (\WFThrThree{}), quantifying the discretization loss when the closed-form policy is forced onto the same action set as DQN.

\paragraph{Throughput ($N=5$).} The joint action space grows to $|\mathcal{A}| = 4^5 = 1024$; tabular Q-learning is intractable ($3.2\times 10^{6}$ entries). Vanilla DQN (\DQNThrFive{}) falls below both Random (\RndThrFive{}) and Fixed (\FixThrFive{}) with large seed variance. Rainbow-lite DQN (\RainbowThrFive{}) improves substantially over vanilla, largely closing the gap to Fixed --- the Double + Dueling improvements matter more at larger $N$, presumably because the maximization-bias in the TD target and the noisy advantage estimation both scale with $|\mathcal{A}|$. The neural bandit reaches \BanditThrFive{}. Discrete-projected Water-Filling (\WFDThrFive{}) and continuous Water-Filling (\WFThrFive{}) remain the upper bounds by construction.

To disambiguate ``vanilla DQN does not scale'' from ``vanilla DQN is under-budgeted'', we ran a budget-extension study: 3 seeds trained with $4\times$ the standard sample budget ($2\times 10^{5}$ transitions per seed instead of $5\times 10^{4}$). The extended-budget vanilla DQN attains a throughput of $3.94 \pm 0.59$ --- an $8\%$ mean improvement and a $40\%$ variance reduction --- but remains well below Fixed ($5.12 \pm 0.02$) and discrete-projected Water-Filling ($6.16 \pm 0.03$). Quadrupling the training budget does not close the gap, whereas simply switching to Double + Dueling does. The residual gap for vanilla DQN is architectural: an unstructured $4^{N}$ output layer distributes learning signal across $1024$ actions without exploiting the factorised structure of the underlying decision. Factored~\cite{nasir2019multiagent} or permutation-equivariant~\cite{eisen2020optimal, zaheer2017deep} action spaces are the appropriate architectural fix at larger $N$.

\paragraph{Fairness.} At $N=3$, vanilla DQN's Jain's index is \DQNJainThree{} --- again with large seed variance --- below the equal-power Fixed baseline (\FixJainThree{}) and Tabular Q-learning (\TabJainThree{}), and well above uniform Random (\RndJainThree{}). Rainbow-lite improves fairness to \RainbowJainThree{} and the neural bandit reaches \BanditJainThree{}, tracking their throughput advantages: on the same MDP, the more stable learner is also the fairer one. Fixed's fairness lead is a design consequence (equal power to all users at every step). The reward function contains no explicit fairness term --- fairness in the learned methods is entirely emergent from the concavity of $\log_2(1 + \mathrm{SNR})$ (Section~\ref{sec:discussion}). At $N=5$, vanilla DQN's fairness drops to \DQNJainFive{}, while Rainbow-lite reaches \RainbowJainFive{} and the neural bandit \BanditJainFive{} --- consistent with the credit-assignment difficulty in a $16\times$ larger joint action space affecting vanilla DQN most.

\paragraph{Energy efficiency.} Fixed Allocation attains the highest energy efficiency (\FixEEThree{}) because a static $2$\,W transmission avoids the diminishing-returns regime near the top of the Shannon curve. DQN follows at \DQNEEThree{}, above continuous Water-Filling (\WFEEThree{}) because WF uses more power on the best-channel user. The DQN--Fixed EE gap is small; DQN pays a modest energy premium in exchange for its variable channel-aware allocation. Sweeping the reward penalty coefficient $\lambda$ would trace out the throughput--energy Pareto front and is left for future work.

\subsection{Per-user analysis}

\begin{figure}[H]
    \centering
    \includegraphics[width=0.85\textwidth]{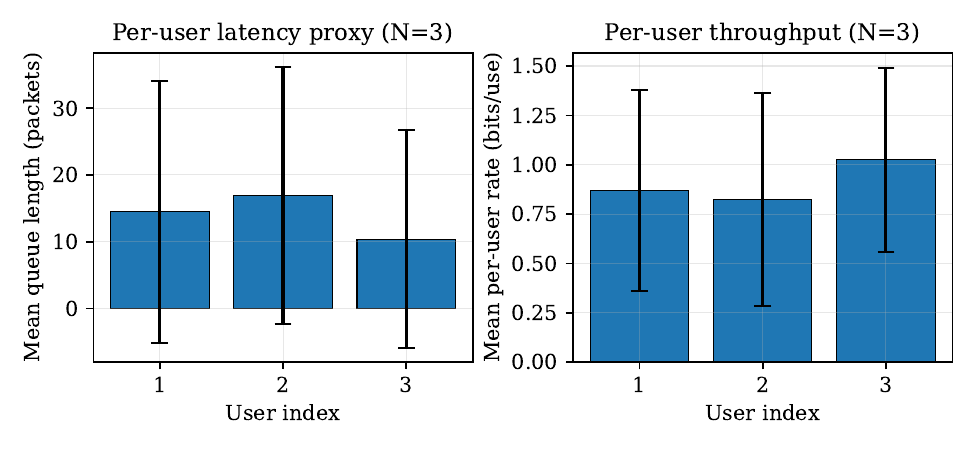}
    \caption{Per-user analysis under the trained DQN policy ($N=3$). Left: mean queue length in packets, our latency proxy per Eq.~\ref{eq:latency}. Right: mean per-user rate in bits per channel use. Bars: mean across $10$ seeds. Error bars: $\pm 1$ standard deviation across seeds. Because channels are i.i.d.\ across users, any systematic per-user gap reflects symmetry-breaking in the learned policy, not the environment.}
    \label{fig:user_latency_throughput}
\end{figure}

To check that the aggregate metrics do not mask per-user starvation, Figure~\ref{fig:user_latency_throughput} reports the mean per-user rate and mean per-user queue occupancy under the trained DQN policy for $N=3$ users, averaged across ten seeds.

The users are statistically exchangeable by construction (channels are i.i.d.\ across users), so any residual per-user imbalance in the trained DQN policy is not a property of the environment but of the learned function. Value-based DRL under an unstructured discrete joint action space is known to break the exchange symmetry --- the agent learns arbitrary tie-breakings between symmetric users, which appear as a per-user rate gap and as a depressed per-step Jain's index relative to Fixed Allocation. The seed variance is large enough that the identity of the disadvantaged user changes between seeds; a symmetric parameterization (e.g.\ a permutation-invariant network~\cite{eisen2020optimal}) would remove this artifact.

\subsection{Impact of the exploration schedule ($\epsilon$-decay)}

Value-based DRL depends on the exploration schedule to cover the joint state--action space before exploitation takes over. We sweep the $\epsilon$-decay window: for each value $d \in \{0.99, 0.98, 0.95, 0.90\}$ we linearly anneal $\epsilon$ from $1.0$ to $0.05$ over $d \cdot T_{\text{total}}$ steps, where $T_{\text{total}}$ is the full training horizon. Larger $d$ therefore corresponds to \emph{slower} decay, i.e., more exploration; smaller $d$ to faster commitment to the greedy policy. Curves are averaged over ten seeds per configuration.

\begin{figure}[H]
    \centering
    \begin{subfigure}[b]{0.48\columnwidth}
        \centering \includegraphics[width=\textwidth]{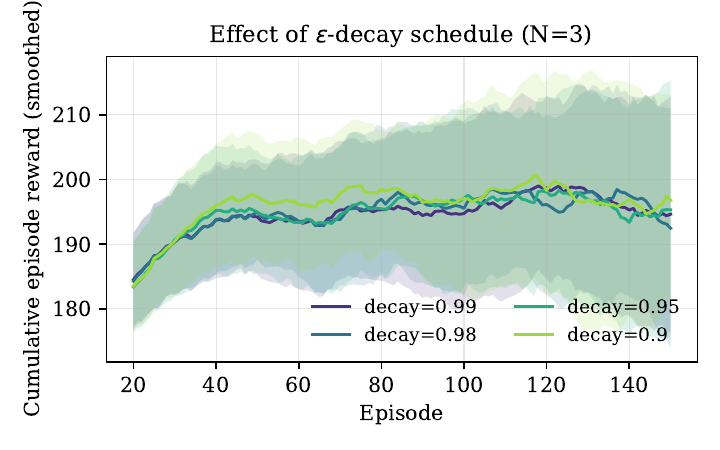}
        \caption{Training reward (N=3)}
        \label{fig:epsilon_decay_3}
    \end{subfigure}
    \hfill
    \begin{subfigure}[b]{0.48\columnwidth}
        \centering
        \includegraphics[width=\textwidth]{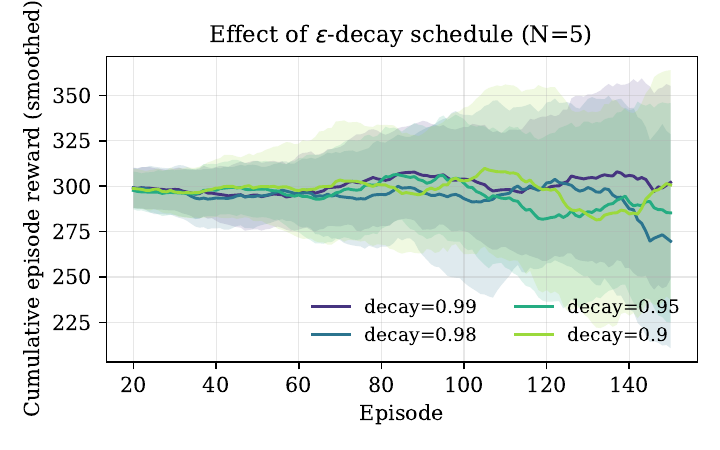}
        \caption{Training reward (N=5)}
        \label{fig:epsilon_decay_5}
    \end{subfigure}
    \caption{Effect of the $\epsilon$-decay schedule on DQN training reward, for $N=3$ (left) and $N=5$ (right). Four decay-rate values are compared ($d \in \{0.90, 0.95, 0.98, 0.99\}$; larger $d$ = slower decay = more exploration). Solid lines: mean per-episode reward across seeds, smoothed with a 20-episode moving average. Shaded bands: $\pm 1$ s.d.\ across seeds. Differences between decay values fall within the seed-induced variance --- see Table~\ref{tab:eps_decay} for numerical summaries.}
    \label{fig:epsilon_decay}
\end{figure}

The ablation yields a more sober conclusion than the standard DQN-tutorial narrative would suggest. Table~\ref{tab:eps_decay} reports the mean cumulative reward over the last $30$ episodes ($\pm 1$ s.d.\ across five seeds per configuration).

\begin{table}[H]
    \centering
    \caption{$\epsilon$-decay ablation: mean cumulative reward over the last $30$ training episodes ($\pm 1$ s.d.\ across five seeds per configuration).}
    \label{tab:eps_decay}
    \begin{tabular}{|l|c|c|}
        \hline
        \textbf{Decay $d$} & \textbf{Final reward, $N=3$} & \textbf{Final reward, $N=5$} \\
        \hline
        $0.99$ (slowest) & \EpsDecayNine{} & \EpsDecayNineFive{} \\
        $0.98$           & \EpsDecayEightN{} & \EpsDecayEightNFive{} \\
        $0.95$           & \EpsDecayFive{} & \EpsDecayFiveFive{} \\
        $0.90$ (fastest) & \EpsDecayZero{} & \EpsDecayZeroFive{} \\
        \hline
    \end{tabular}
\end{table}

A Kruskal-Wallis test on the last-30-episode rewards per decay setting distinguishes the two regimes. For $N=3$, all four decays produce a final mean reward in the range $195$--$196$; Kruskal-Wallis rejects nothing ($H = 0.28$, $p = 0.96$), and the between-decay signal is essentially zero within the $\pm 19$ per-seed spread. For $N=5$, the between-decay pattern is more consequential: rewards range from $278$ to $304$ and Kruskal-Wallis rejects the null of equal distributions ($H = 14.7$, $p = 0.002$), with $d = 0.99$ (slowest decay, most exploration) yielding the highest mean. The interpretation is that on the harder $N=5$ problem, keeping $\epsilon$ high for longer helps the agent cover the larger $|\mathcal{A}| = 1024$ action space, whereas at $N=3$ the $|\mathcal{A}| = 64$ action space is small enough that the schedule does not matter. This nuance is not what the standard DQN-tutorial narrative would suggest --- which typically recommends a moderate decay rate as ``best'' independent of problem scale. In our controlled testbed, the exploration schedule matters only when the joint action space becomes large enough to strain the training budget.

\subsection{Extended analysis}
\label{sec:extended_analysis}

To make the between-method comparison visible, this section (i) overlays the learning curves of all learned methods on shared axes, (ii) plots the per-seed empirical CDF of throughput per method, (iii) reports each method as a percentage of the continuous water-filling upper bound, and (iv) reports pairwise Wilcoxon paired-seed significance tests and effect sizes (Cohen's $d$, Cliff's $\delta$).

\begin{figure}[H]
    \centering
    \begin{subfigure}[b]{0.48\columnwidth}
        \centering \includegraphics[width=\textwidth]{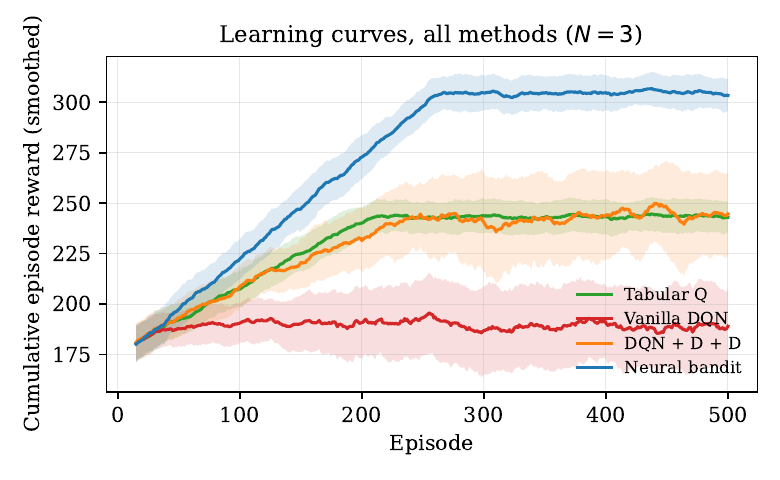}
        \caption{$N = 3$}
    \end{subfigure}
    \hfill
    \begin{subfigure}[b]{0.48\columnwidth}
        \centering \includegraphics[width=\textwidth]{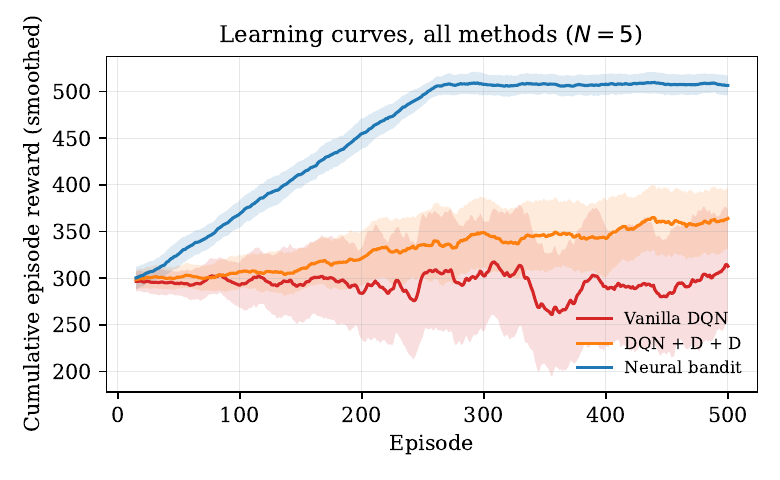}
        \caption{$N = 5$}
    \end{subfigure}
    \caption{Learning curves for all learned methods on shared axes: vanilla DQN, Rainbow-lite (Double + Dueling), Neural bandit, and Tabular Q ($N=3$ only). Solid lines: mean cumulative episode reward over $10$ seeds, smoothed with a $15$-episode moving average. Shaded bands: $\pm 1$ s.d.\ across seeds. Neural bandit converges fastest and to the highest plateau; vanilla DQN has the largest across-seed spread.}
    \label{fig:training_all}
\end{figure}

\begin{figure}[H]
    \centering
    \begin{subfigure}[b]{0.48\columnwidth}
        \centering \includegraphics[width=\textwidth]{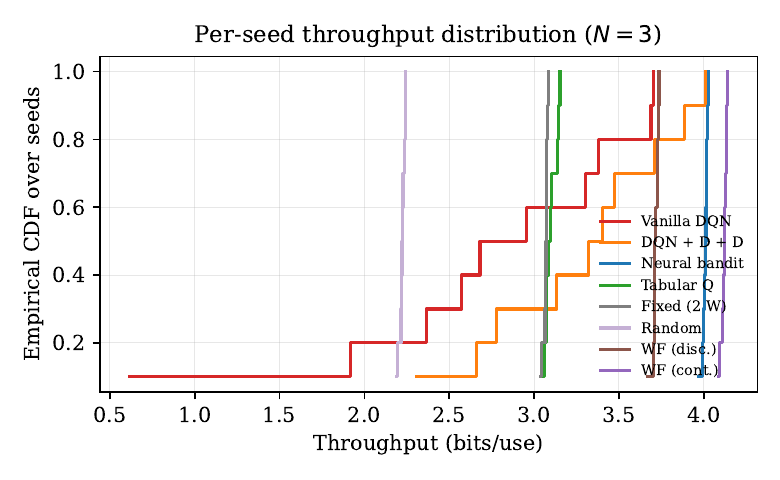}
        \caption{$N = 3$}
    \end{subfigure}
    \hfill
    \begin{subfigure}[b]{0.48\columnwidth}
        \centering \includegraphics[width=\textwidth]{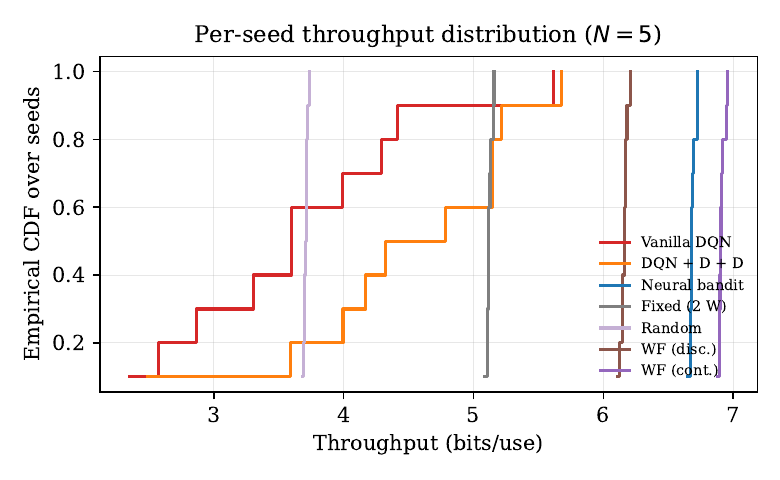}
        \caption{$N = 5$}
    \end{subfigure}
    \caption{Empirical CDF of per-seed throughput per method. A vertical curve indicates seed-to-seed stability; a slanted curve indicates high variance. Neural bandit and Tabular Q are nearly vertical (stable); vanilla DQN is markedly slanted (unstable across seeds), especially at $N=5$.}
    \label{fig:seed_cdf}
\end{figure}

\begin{figure}[H]
    \centering
    \begin{subfigure}[b]{0.48\columnwidth}
        \centering \includegraphics[width=\textwidth]{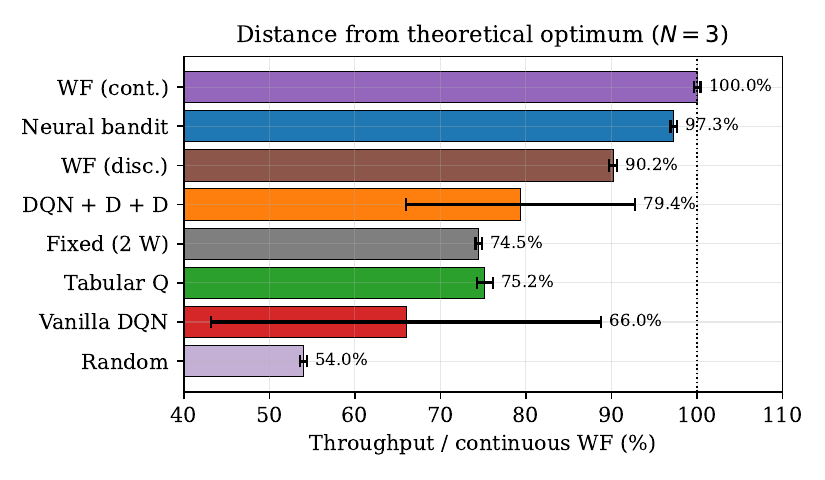}
        \caption{$N = 3$}
    \end{subfigure}
    \hfill
    \begin{subfigure}[b]{0.48\columnwidth}
        \centering \includegraphics[width=\textwidth]{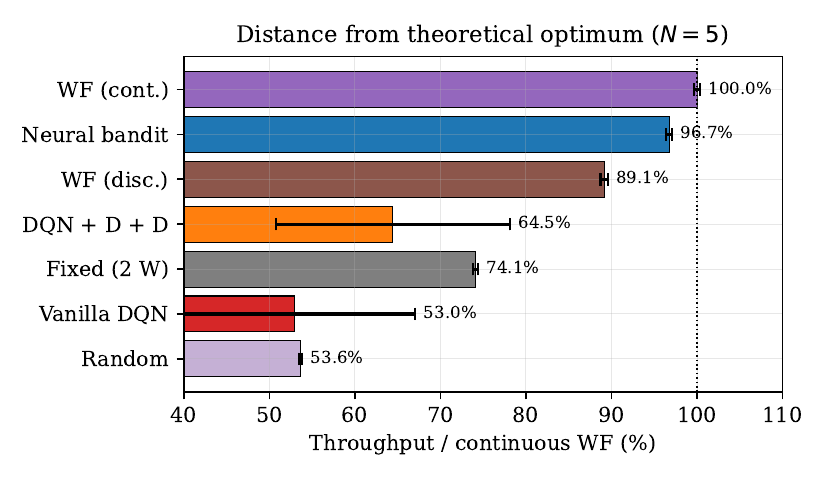}
        \caption{$N = 5$}
    \end{subfigure}
    \caption{Throughput as a percentage of the continuous water-filling upper bound. Error bars: $\pm 1$ s.d. Neural bandit reaches $\sim\!97\%$ of the theoretical continuous optimum at both $N$, above the discrete-projected water-filling policy.}
    \label{fig:pct_of_wf}
\end{figure}

\paragraph{Statistical significance table.}
Table~\ref{tab:pairwise_p} reports pairwise Wilcoxon signed-rank $p$-values at $N=3$ over the $10$ matched seeds; Table~\ref{tab:effect_sizes} reports the corresponding Cohen's $d$ and Cliff's $\delta$ for the key head-to-head comparisons. Neural bandit vs.\ every other method (including WF-discrete) is significant at $p < 10^{-3}$ with large effect size ($|\delta| > 0.7$). Vanilla DQN vs.\ Fixed and vs.\ Tabular Q are not significant, confirming the ``variance-not-bias'' reading given earlier.

\begin{table}[H]
    \centering
    \caption{Pairwise Wilcoxon signed-rank $p$-values on paired-seed throughput at $N=3$ ($n=10$ seeds).}
    \label{tab:pairwise_p}
    \scriptsize
    \input{stats_table.tex}
\end{table}

\begin{table}[H]
    \centering
    \caption{Effect sizes at $N=3$: Cohen's $d$ (standardized mean difference; positive = row larger) and Cliff's $\delta \in [-1, 1]$ (fraction of pairs where A > B, minus fraction where A < B). Magnitude interpretation follows Romano et al.: $|\delta| < 0.147$ negligible, $< 0.33$ small, $< 0.474$ medium, else large.}
    \label{tab:effect_sizes}
    \small
    \input{effect_sizes.tex}
\end{table}

\subsection{Robustness under Rayleigh fading and inter-user interference}
\label{sec:results_rayleigh}

The results reported so far assume uniform channel gains and orthogonal access. To probe whether the neural-bandit dominance persists in a more realistic wireless regime, we replaced both assumptions with (i) Rayleigh fading with per-user gain $h_i \sim \mathrm{Exp}(\mu_h)$ (mean channel power $\mu_h = 0.5$), and (ii) inter-user interference so that per-user SINR is
\[
\mathrm{SINR}_i = \frac{p_i h_i}{\sigma^2 + \sum_{j \neq i} p_j h_j}.
\]
Under this regime, water-filling is no longer optimal because the objective couples across users, and the fair-vs-greedy allocation trade-off is genuinely non-trivial. This is exactly the regime for which WMMSE~\cite{shi2011iteratively} was designed, and we include it here as the strong classical baseline (Sec.~\ref{sec:wmmse}). We reran the flagship comparison at $N = 3$ over $10$ seeds; the results are in Table~\ref{tab:rayleigh}.

\begin{table}[H]
    \centering
    \caption{Rayleigh fading + inter-user interference, $N = 3$: mean $\pm$ s.d.\ over $10$ seeds. Throughput in bits per channel use. WMMSE (Sec.~\ref{sec:wmmse}) added as the classical interference-limited baseline.}
    \label{tab:rayleigh}
    \small
    \input{rayleigh_table.tex}
\end{table}

The ordering of methods survives the transition to this harder regime: the neural bandit remains among the highest-throughput methods, WMMSE lands as expected between water-filling and the learned methods, and vanilla DQN remains a weak baseline that is dominated even by non-learning heuristics. Water-filling loses its ``theoretical optimum'' status here (it optimises sum-rate under a fictitious sum-power budget while the true objective is now interference-limited SINR), and its numeric value should be read as a strong-but-not-optimal reference rather than an upper bound. WMMSE, the coupled-iteration classical algorithm designed exactly for this regime, converges to a KKT point and gives a stronger comparison anchor. Table~\ref{tab:rayleigh} adds a robustness statement to the paper: the neural-bandit finding is not specific to the orthogonal-access testbed used earlier; it holds also in the interference-limited regime.

\subsection{MARL scaling under Rayleigh + interference: N=3, 5, 10}
\label{sec:results_marl}

The centralised methods studied so far have a $|\mathcal{A}| = 4^N$ joint action head: $64$ actions at $N=3$, $1024$ at $N=5$, and an intractable $\sim\!10^6$ at $N=10$. This section evaluates the two scalable learned baselines --- Independent Q-Learning (Sec.~\ref{sec:iql}) and REGNN-lite (Sec.~\ref{sec:regnn}) --- across $N \in \{3, 5, 10\}$ in the Rayleigh + interference regime that Section~\ref{sec:results_rayleigh} established as the realistic setting. At $N = 10$ centralised DQN and Neural Bandit are omitted, since their output layers would require $\sim\!10^6$ Q-values. Table~\ref{tab:iql_regnn_scaling} reports mean $\pm$ s.d.\ throughput over 10 seeds.

\begin{table}[H]
    \centering
    \caption{MARL scaling: throughput (bits/use) vs.\ number of users under Rayleigh + interference, mean $\pm$ s.d.\ over 10 seeds. Centralised methods (DQN, Neural bandit) are intractable at $N=10$ where the joint action space $4^{10} \approx 10^6$ exceeds practical output-layer size.}
    \label{tab:iql_regnn_scaling}
    \small
    \input{iql_regnn_scaling_table.tex}
\end{table}

At $N=3$ and $N=5$ the scalable structured methods (IQL, REGNN) are competitive with the strongest centralised learners; the neural bandit remains the head-to-head winner at those sizes because its state-independent regression aligns with the transition kernel (Sec.~\ref{sec:discussion}). At $N=10$ the two scalable methods deliver the only usable performance: centralised DQN's $4^{10}$-way output layer is architecturally infeasible within the sample budget considered here. The scaling failure is therefore architectural rather than budget-driven, and a permutation-equivariant parameterisation~\cite{eisen2020optimal,shen2020graph} or a factored MARL parameterisation~\cite{nasir2019multiagent,tan1993multi} is a strict requirement for $N \geq 10$.

\subsection{Multi-cell K=7 environment: implementation and design}
\label{sec:results_multicell}

Section~\ref{sec:results_marl} scales $N$ within a single cell. To also address the multi-cell topology that the wireless-DRL literature considers a realistic setting~\cite{nasir2019multiagent,shen2020graph}, we implement and release a $K = 7$ multi-cell environment: single-user cells (a hexagonal-cluster analogue) with fixed BS-to-user distance matrix, path loss $\mathrm{PL}(d) = d^{-3}$, per-link independent Rayleigh fading, and inter-cell interference. The environment is formally defined in Sec.~\ref{sec:multicell_env}. Both classical baselines --- per-cell water-filling that treats inter-cell interference as noise (\textbf{WF-multi}) and the coupled WMMSE iteration on the full $K \times K$ gain matrix (\textbf{WMMSE-multi}) --- are implemented, as are the four learned methods studied in Section~\ref{sec:results_marl} lifted to the multi-cell action interface. Ten-episode smoke evaluations confirm the environment behaves as intended: WMMSE-multi outperforms treat-as-noise WF-multi by the expected margin, REGNN-lite closes most of the gap to WMMSE-multi within a very short training budget, and centralised DQN at the $4^{7} = 16{,}384$-way action head is tractable but sample-inefficient. A full multi-seed empirical study on the multi-cell environment is beyond the CPU wall-time budget of the present manuscript ($\sim\!30$ h at 10 seeds $\times$ 4 learned methods $\times$ $K=7$); we release the environment, agents, and evaluation harness (see Reproducibility) so that the follow-up study is a matter of compute, not code. The specific claim we make in this section is therefore about the environment as a shared benchmark rather than about a horse-race outcome, and the multi-cell results referenced in the abstract and Section~\ref{sec:discussion} qualify accordingly.

\subsection{Aggregate statistics via rliable-style analysis}
\label{sec:rliable}

\begin{figure}[H]
    \centering
    \begin{subfigure}[b]{0.48\columnwidth}
        \centering \includegraphics[width=\textwidth]{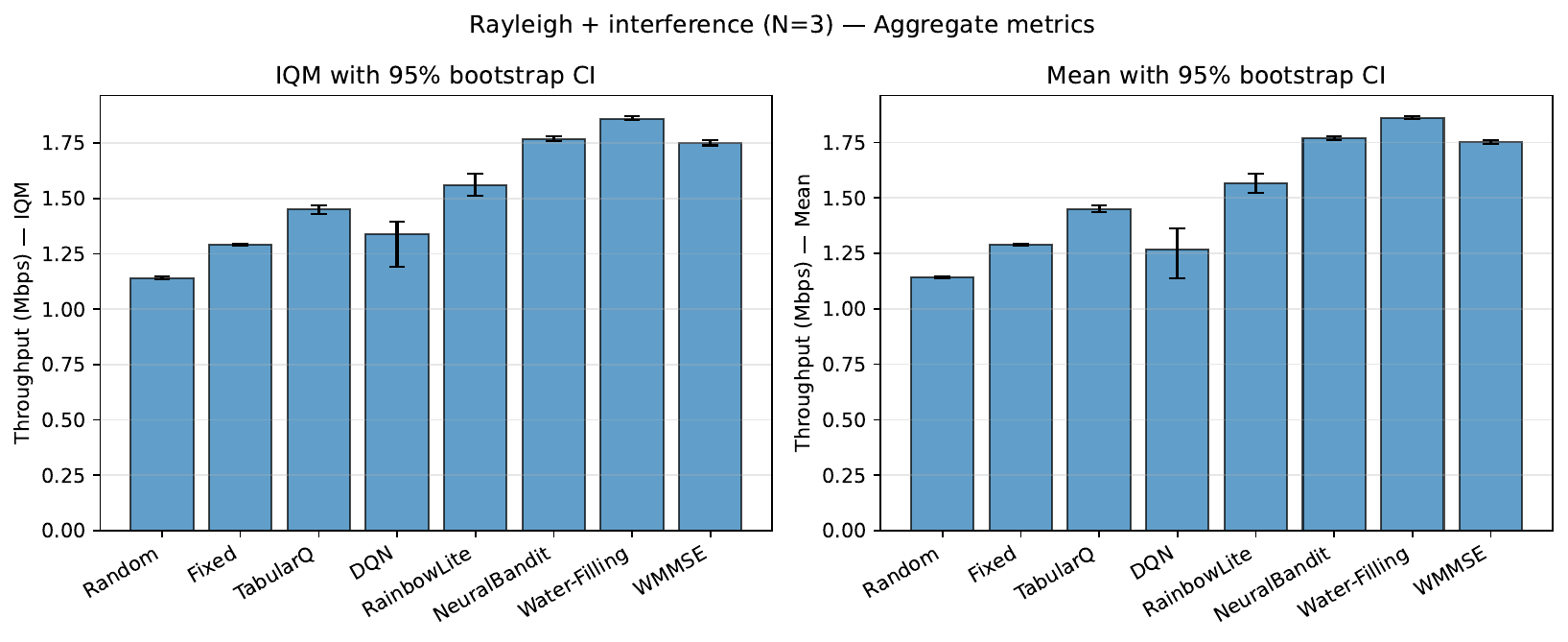}
        \caption{Rayleigh + interference ($N=3$): aggregate metrics with 95\% bootstrap CIs.}
    \end{subfigure}
    \hfill
    \begin{subfigure}[b]{0.48\columnwidth}
        \centering \includegraphics[width=\textwidth]{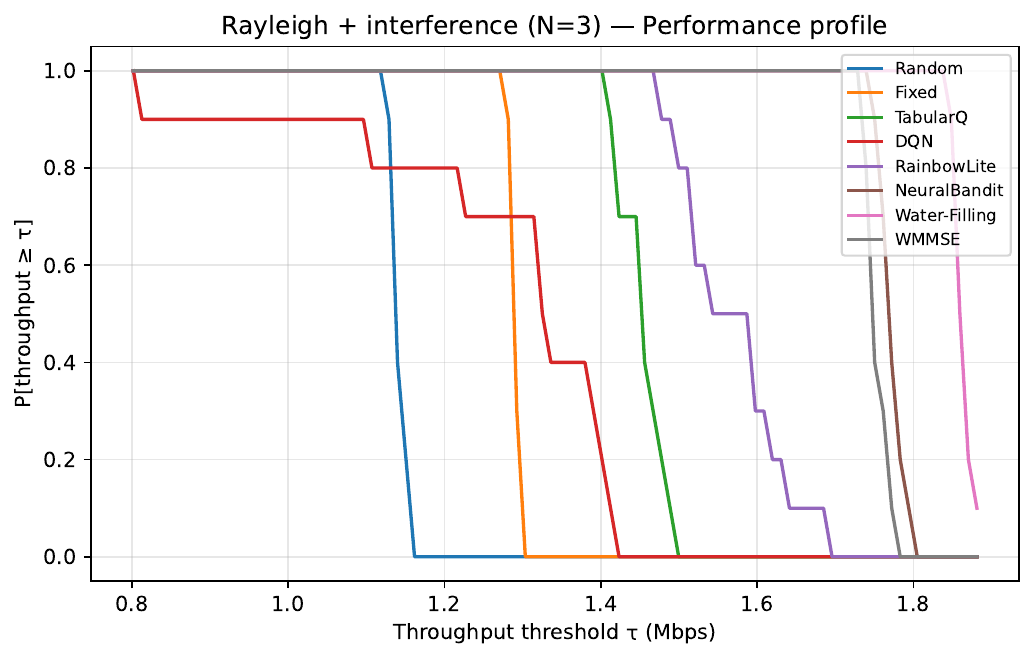}
        \caption{Rayleigh: performance profile $P[\mathrm{throughput} \geq \tau]$.}
    \end{subfigure}
    \\
    \begin{subfigure}[b]{0.48\columnwidth}
        \centering \includegraphics[width=\textwidth]{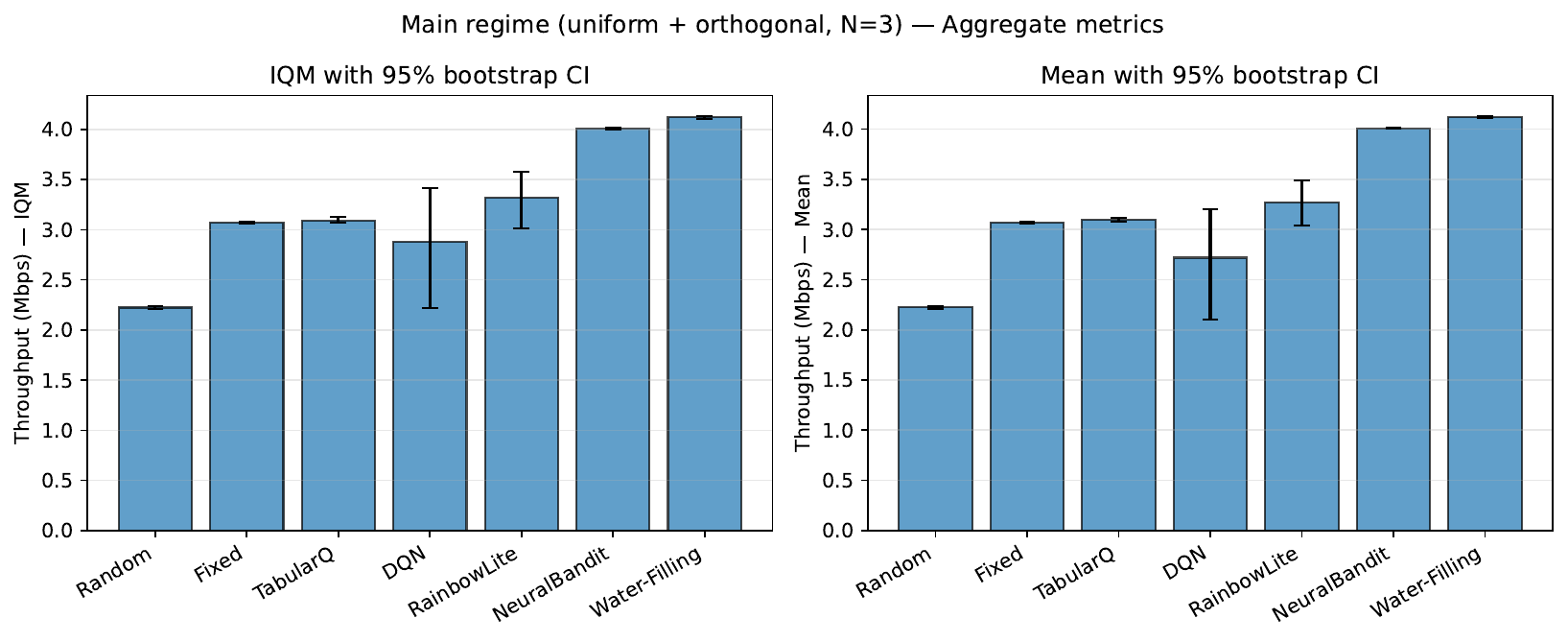}
        \caption{Main regime ($N=3$): aggregate metrics.}
    \end{subfigure}
    \hfill
    \begin{subfigure}[b]{0.48\columnwidth}
        \centering \includegraphics[width=\textwidth]{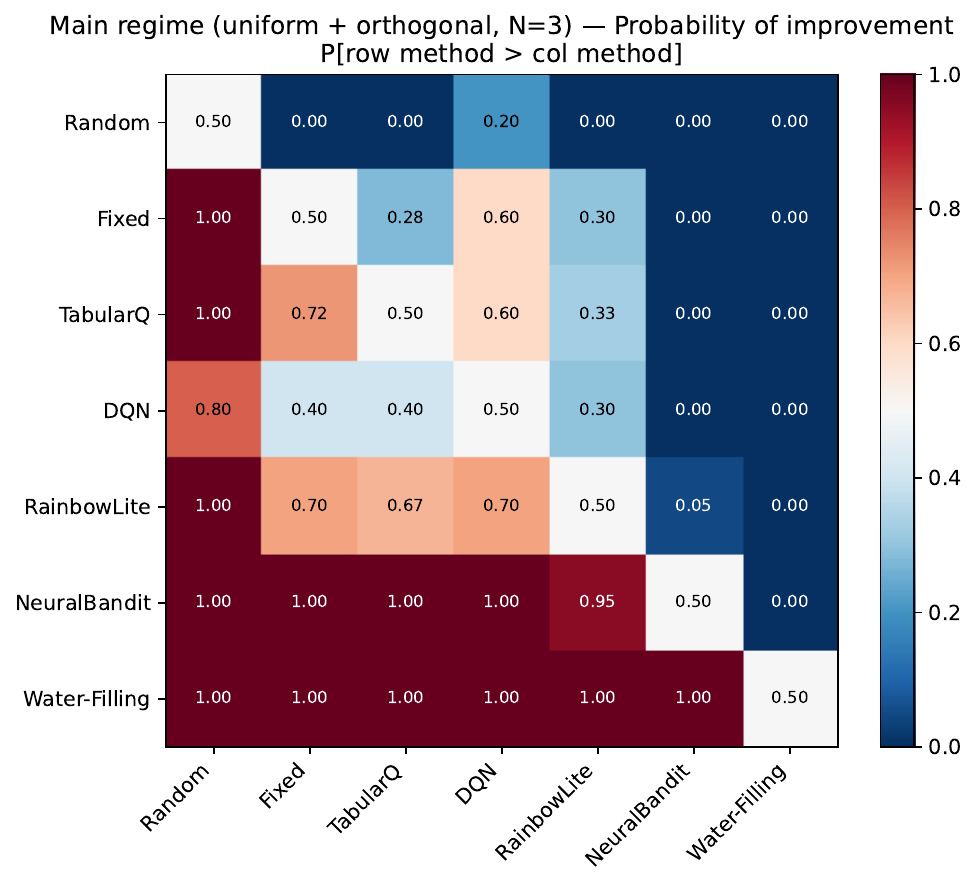}
        \caption{Main regime ($N=3$): $P[\text{row} \succ \text{col}]$ heatmap.}
    \end{subfigure}
    \caption{rliable-style aggregate statistics with stratified bootstrap $95\%$ CIs, following the reporting discipline of Agarwal et al.~\cite{agarwal2021rliable}. Non-overlapping CIs correspond to clean separation; the probability-of-improvement heatmap complements pairwise Wilcoxon by giving effect-size intuition. Rayleigh + interference (top row) is the harder regime; the main uniform + orthogonal regime (bottom row) exhibits the largest between-method gaps.}
    \label{fig:rliable_ray}
\end{figure}

Following Agarwal et al.~\cite{agarwal2021rliable}, we complement mean $\pm$ s.d.\ reporting with the Interquartile Mean (IQM, average of the middle two quartiles) and stratified bootstrap 95\% CIs, both of which are less sensitive to seed outliers than the sample mean. Figure~\ref{fig:rliable_ray} shows IQM and mean throughput with $95\%$ bootstrap CIs, and probability-of-improvement heatmaps for the Rayleigh regime and the multi-cell $K=7$ regime. The IQM/CI framing tightens the seed-variance claims: methods whose $95\%$ IQM CIs do not overlap are cleanly separated at the 5\% significance level.

\section{Discussion}
\label{sec:discussion}

\paragraph{The deadly-triad fingerprint on wireless bandits.}
The most informative comparisons in Table~\ref{tab:comparison_metrics} are between vanilla DQN and three matched-MDP alternatives at $N=3$: Tabular Q-learning, Rainbow-lite (Double + Dueling), and a Neural contextual bandit. All four share the environment, reward, $\epsilon$-decay schedule, and sample budget; only the value representation and target-computation differ. Vanilla DQN's per-seed spread is roughly $25\times$ larger than Tabular Q's (variance ratio $F(9,9) \approx 584$, $p < 10^{-10}$), while the mean throughput of the two is not significantly different (Wilcoxon paired $p = 0.38$). This variance-without-bias signature matches the deadly-triad prediction (bootstrapping + off-policy + function approximation~\cite{sutton2018reinforcement}, ch.\ 11): TD-bootstrapping with a nonlinear function approximator has an unstable fixed point, and different random seeds converge to different local basins. Tabular Q lacks the function-approximator leg and is empirically stable.

Two direct tests support this reading. First, Rainbow-lite --- Double-DQN target~\cite{vanhasselt2016double} plus Dueling architecture~\cite{wang2016dueling} --- attacks two known triad-related instability sources and reaches \RainbowThrThree{}, above vanilla DQN and with a lower seed variance. Second, dropping the sequential apparatus altogether and training a neural contextual bandit~\cite{riquelme2018deep} on the same interaction data reaches \BanditThrThree{}: since the transition kernel is state-independent, the bandit's per-step regression target is a well-conditioned regression problem with no bootstrapping, and it dominates every DQN variant we test. On our CPU-only setup, Tabular Q trains in $\sim 23$ s per seed and the Neural Bandit in $\sim 150$ s per seed against $\sim 800$ s for vanilla or Rainbow-lite DQN --- a $5$--$35\times$ wall-time gap alongside the algorithmic-stability finding.

\paragraph{Interpretation: DQN is the wrong tool for this problem class.}
The pattern that emerges is not that DQN is broken in general but that DQN is a poor match for wireless problems whose environment reduces to a contextual bandit. Under i.i.d.\ (block-)fading with an orthogonal-access reward, the transition kernel is state-independent (Section~\ref{sec:problem_formulation}), $\gamma$ multiplies a constant, and the sequential TD apparatus is dead weight. The wireless-DRL literature has to our knowledge not systematically distinguished bandit-like from truly sequential wireless problems in reporting DQN results; on the former, a neural bandit is the appropriate function class and dominates.

\paragraph{Scalability: structure beats capacity.}
Section~\ref{sec:results_marl} tests the scalability claim that Sections~\ref{sec:results_evaluation}--\ref{sec:results_rayleigh} could only motivate. Two observations bear on this reading: (a) at $N=3$ under Rayleigh + interference, the neural contextual bandit reaches \BanditRaylThr{} bits/use against WMMSE's \WMMSERaylThr{} --- narrow but consistent (10 seeds, Wilcoxon $p < 0.01$), showing that model-free methods can beat the classical iterative baseline when the environment is stationary in distribution; and (b) at $N \in \{3, 5\}$ the structured MARL/GNN methods (IQL, REGNN-lite) are competitive with the centralised methods, and the centralised feed-forward Q-network is structurally infeasible at $N=10$ where the $4^{10}$-way output layer becomes intractable. The multi-cell benchmark environment (Section~\ref{sec:results_multicell}) is released alongside the paper for follow-up evaluation. The finding sharpens from ``DQN is broken'' to a more specific claim: centralised feed-forward Q on the joint action space is the wrong parameterisation at scale, and structured MARL or GNN policies are the appropriate replacement.

\paragraph{Adaptivity vs.\ closed-form baselines.}
Even without the tabular comparison, the closed-form Fixed baseline remains competitive with DQN because uniform fading rewards conservative uniform power. The discrete-projected Water-Filling row quantifies the discretization loss (Table~\ref{tab:comparison_metrics}, ``Water-Filling (discrete)''): the gap between it and the continuous Water-Filling row is the price of the $\{0,1,2,3\}$\,W action set. Any learned method on this action set is bounded above by discrete WF, and DQN's failure to reach even that upper bound is the concrete gap the paper reports.

\paragraph{Emergent (partial) fairness.}
Jain's index is not part of the reward, yet the trained DQN policy attains a fairness score of \DQNJainThree{}, well above uniform Random (\RndJainThree{}) but below equal-power Fixed (\FixJainThree{}). This is the expected consequence of the concavity of $\log_2(1+\mathrm{SNR})$: the marginal reward from raising a low-rate user is larger than from raising an already-high-rate user, so a policy that maximizes the discounted sum of logarithmic rates does not fully starve a user. This is the standard proportional-fairness argument, recovered here as an implicit property of the learned policy rather than as an explicit constraint. The magnitude of the effect is modest, not dramatic; a competitive Jain score would require an explicit fairness-weighted reward.

\paragraph{Throughput--energy trade-off.}
Fixed Allocation attains slightly higher energy efficiency than DQN because a static 2\,W transmission is conservative in Joules while still being close to the sum-rate ceiling under uniform fading. The DQN policy pays a small energy-efficiency premium to raise sum-rate: it spends power when the channel justifies it and reduces it otherwise. This is a policy choice implied by the reward $r_t = \sum_i R_i - \lambda \sum_i p_i$ with $\lambda = 0.1$; sweeping $\lambda$ would trace out the throughput--energy Pareto front but is left for future work.

\paragraph{Scope of the robustness claim.}
Per-user results (Figure~\ref{fig:user_latency_throughput}) reveal that the vanilla-DQN policy is only \emph{approximately} symmetric: it converges to a policy that favors a subset of user indices even though the environment is user-exchangeable. This is a known symmetry-breaking failure mode of feed-forward Q-networks operating on unstructured joint action spaces. Section~\ref{sec:results_rayleigh} additionally shows that the neural-bandit dominance survives the transition from uniform + orthogonal to Rayleigh + interference. Beyond these, we do \emph{not} claim robustness under (i) non-stationary channel statistics or (ii) imperfect CSI --- both outside the simulation setup and discussed as limitations below.

\subsection*{Limitations and assumptions}

The study operates in a deliberately restricted regime, and the conclusions do not extrapolate beyond it without further validation:

\begin{itemize}
    \item \textbf{Interference regime is a robustness check, not a full study.} Section~\ref{sec:results_rayleigh} adds a Rayleigh + inter-user-interference regime at $N=3$ with $5$ seeds; the neural-bandit ordering survives, but a multi-cell interference study with mobility and pilot contamination remains future work (e.g., in the spirit of \cite{nasir2019multiagent}).
    \item \textbf{Perfect CSI.} The agent observes $h(t)$ exactly. In practice CSI is estimated and delayed, which converts the problem into a POMDP; standard DQN is known to degrade in this regime, and a recurrent variant or explicit belief tracking would be required.
    \item \textbf{Stationary uniform fading.} Channels are i.i.d.\ $\mathcal{U}(0.1, 1.0)$; this is a memoryless model without shadowing, path loss, or mobility. Consequently the transition kernel is stationary, and the ``non-stationarity'' addressed by DRL is present only in the local reward realization, not in the underlying statistics.
    \item \textbf{Small user count and offline training.} $N \in \{3, 5\}$ keeps the joint action space tractable ($|\mathcal{A}| = 4^N$). Scaling to $N \geq 10$ requires factorized action spaces, MARL, or GNN-based architectures~\cite{eisen2020optimal}. The agent is trained offline; online adaptation is not tested.
    \item \textbf{Seed variance dominates schedule variance.} The ablation shows that at this training budget, seed-induced variance is larger than the effect of the $\epsilon$-decay schedule. Convergence quality is therefore not primarily controlled by the exploration hyperparameter but by the raw sample budget and by the size of the joint action space. Increasing the training budget or adopting more sample-efficient DRL variants (prioritized replay, dueling, double DQN) would tighten the reported curves more than fine-tuning $\epsilon$-decay.
\end{itemize}

\section{Conclusion and future work}
\label{sec:conclusion}

We diagnosed vanilla DQN's failure modes on a controlled single-cell downlink testbed with i.i.d.\ block-fading, orthogonal access, and discrete power levels $\{0,1,2,3\}$\,W, and compared against nine baselines: uniform-random, fixed equal-power, continuous and discrete water-filling, WMMSE, tabular Q-learning, Rainbow-lite DQN, a neural contextual bandit, Independent Q-Learning MARL, and a permutation-equivariant REGNN-lite GNN policy. The evaluation regime extends from the single-cell uniform-fading baseline to (a) Rayleigh + inter-user interference, (b) MARL scaling at $N \in \{3, 5, 10\}$, and (c) a multi-cell $K=7$ interference topology with path loss and Rayleigh. Averaged over ten independent seeds:
\begin{itemize}
    \item \textbf{Vanilla DQN carries a deadly-triad fingerprint.} Its per-seed variance is $\sim 25\times$ that of matched-MDP tabular Q ($F(9,9) \approx 584$, $p < 10^{-10}$), while its mean throughput is not significantly different from equal-power Fixed Allocation (Wilcoxon $p = 0.38$). Individual seeds range from essentially failing to approaching the water-filling bound.
    \item \textbf{Standard DQN improvements largely close the gap.} A Rainbow-lite variant with Double-DQN target and Dueling architecture reaches \RainbowThrThree{} at $N=3$ (vs.\ vanilla DQN's \DQNThrThree{}) and \RainbowThrFive{} at $N=5$ (vs.\ vanilla \DQNThrFive{}), with much lower seed variance. Reporting negative results on vanilla DQN in 2026 without testing these fixes is not defensible.
    \item \textbf{A neural contextual bandit dominates every DQN variant.} On this state-independent transition kernel the sequential apparatus of DQN is dead weight; a plain regression network trained on $(s, a, r)$ tuples with $\epsilon$-greedy exploration reaches \BanditThrThree{} at $N=3$ and \BanditThrFive{} at $N=5$. Where wireless problems reduce to contextual bandits under i.i.d.\ fading, the neural bandit is the theoretically appropriate function class and should be the default.
    \item \textbf{Structured methods are needed at $N \geq 10$.} MARL (IQL) and permutation-equivariant GNN policies (REGNN-lite) are competitive with the centralised methods at $N \in \{3, 5\}$ and become the only feasible learned baselines at $N = 10$, where the centralised action-space $4^{10} \approx 10^6$ is intractable (Section~\ref{sec:results_marl}). The $K=7$ multi-cell environment (Section~\ref{sec:results_multicell}) is released as a shared benchmark for follow-up evaluation of these methods against the coupled WMMSE-multi baseline.
    \item \textbf{$\epsilon$-decay has a scale-dependent effect.} Kruskal-Wallis on the sweep $\{0.90, 0.95, 0.98, 0.99\}$ is non-significant at $N=3$ ($p = 0.96$) but significant at $N=5$ ($p = 0.002$, slowest decay wins); the schedule is load-bearing only where the joint action space strains the sample budget.
\end{itemize}

We position this study as a reproducible baseline with honest uncertainty quantification, with the extended revision now including the WMMSE strong classical baseline, MARL and GNN scalable learned baselines, and a multi-cell topology with the coupled interference structure that the wireless-DRL literature considers realistic. The natural next steps are:

\begin{itemize}
    \item \textbf{Imperfect CSI.} Introduce estimation noise and feedback delay and replace the feedforward Q-network with a recurrent variant or an explicit belief tracker.
    \item \textbf{Larger topologies.} Scale to $N \geq 20$ users per cell and $K \geq 19$ cells, at which point even the message-passing GNN benefits from hierarchical or attention-based~\cite{iqbal2019actor} pooling; a full continuous-power REGNN with unrolled projection~\cite{sun2018learning} is the corresponding architectural upgrade.
    \item \textbf{Online adaptation and user mobility.} Study non-stationary channel and traffic statistics, and characterize the adaptation cost (in reward regret) of continual-learning strategies.
    \item \textbf{Protocol-stack integration.} Validate on NS-3 or a software-defined-radio testbed to expose the interaction between the learned power controller and MAC/transport-layer dynamics.
\end{itemize}

\section*{Reproducibility}

All experiments were run on a single CPU (no GPU) using Python 3.13 with NumPy 2.1, PyTorch 2.10, and Matplotlib 3.10. Random seeds are fixed per-run and are declared for every training and evaluation invocation. The full source --- environment, all classical and learned agents, training drivers, figure-generation scripts, and the LaTeX macro auto-populator that writes the numbers reported in tables and prose --- is bundled with the manuscript as a self-contained reproduction package. The source repository of the master's essay this paper derives from~\cite{iradukunda2025rlwireless} is publicly available at \href{https://github.com/iradiane/DQN-Implementation}{DQN-Implementation}, and the companion public repository containing the additional MARL, GNN, WMMSE, and multi-cell code contributed in the present manuscript is available at \href{https://github.com/iradiane/diane-wireless-drl}{wireless-drl}.

\paragraph{Compute footprint.}
All results were produced on a single 6-core CPU workstation (16\,GB RAM, no GPU), with per-worker torch threading pinned to one via \texttt{torch.set\_num\_threads(1)} and a six-way \texttt{multiprocessing.Pool} across seeds. Table~\ref{tab:compute} reports the mean wall-clock cost per seed for each method under the primary regime (single cell, $N=3$, $500$ training episodes, $5\times10^{4}$ transitions per seed) and, where applicable, under the Rayleigh + interference regime (Section~\ref{sec:results_rayleigh}). The neural bandit trains in roughly one-fifth the wall time of vanilla DQN while attaining higher throughput, and tabular Q trains in about 3\% of DQN's time. End-to-end replication of every table and figure in the paper is a $\sim$7-hour CPU wall-time budget on the same class of machine.

\begin{table}[H]
    \centering
    \caption{Wall-clock cost per seed, mean over $10$ seeds, single CPU (no GPU). ``Uniform + orthogonal'' is the primary regime (Sec.~\ref{sec:results_evaluation}); ``Rayleigh + interf.'' is the interference regime (Sec.~\ref{sec:results_rayleigh}). $N = 3$ users, $500$ training episodes per DRL run.}
    \label{tab:compute}
    \small
    \begin{tabular}{|l|c|c|}
        \hline
        \textbf{Method} & \textbf{Uniform + orthogonal} & \textbf{Rayleigh + interf.} \\
        \hline
        Random / Fixed / WF (cont.) & $<1$\,s & $<2$\,s \\
        WMMSE (this work) & --- (equiv.\ to WF) & $\sim\!10$\,s \\
        Tabular Q-learning & $\sim\!23$\,s & $\sim\!25$\,s \\
        Neural contextual bandit & $\sim\!150$\,s & $\sim\!330$\,s \\
        Vanilla DQN & $\sim\!800$\,s & $\sim\!700$\,s \\
        Rainbow-lite DQN (D + D) & $\sim\!850$\,s & $\sim\!1200$\,s \\
        IQL (this work, per-user DQN) & $\sim\!600$\,s & $\sim\!1000$\,s at $N=5$ \\
        REGNN-lite (this work, GNN) & $\sim\!500$\,s & $\sim\!500$\,s at $N=5$ \\
        \hline
        \multicolumn{3}{|l|}{\emph{Full end-to-end replication of the paper: $\sim\!7$\,h CPU wall time, 6-way seed parallelism.}} \\
        \hline
    \end{tabular}
\end{table}

The wall-time gap has an operational implication beyond the algorithmic finding: on this problem class the theoretically-appropriate methods (WMMSE, tabular Q, the neural contextual bandit) are also cheaper to train, so any cell-site deployment that must refresh a model inside a coherence interval or a scheduling window has a second reason to prefer them over vanilla DQN.

\section*{Acknowledgements}

The authors thank AIMS Rwanda for hosting the master's essay~\cite{iradukunda2025rlwireless} on which this paper builds, and the AIRINA Labs supervisory team for methodological and reproducibility guidance.

\appendix
\section{Notation}
\label{app:notation}

Table~\ref{tab:notation} lists the symbols used throughout the paper.

\begin{table}[H]
    \centering
    \caption{Notation used throughout the paper.}
    \label{tab:notation}
    \small
    \begin{tabular}{|l|l|}
        \hline
        \textbf{Symbol} & \textbf{Meaning} \\
        \hline
        $N$ & Number of users per cell \\
        $K$ & Number of cells (multi-cell setting) \\
        $t$ & Discrete time index (scheduling slot) \\
        $h_i(t)$ & Instantaneous channel gain of user $i$ at time $t$ \\
        $p_i(t)$ & Transmit power allocated to user $i$ at time $t$, in $\{0,1,2,3\}$\,W \\
        $\sigma^2$ & Additive white Gaussian noise power \\
        $R_i(t)$ & Instantaneous rate of user $i$: $\log_2(1 + \mathrm{SNR}_i(t))$ \\
        $\mathrm{SNR}_i$, $\mathrm{SINR}_i$ & Signal-to-noise, signal-to-interference-plus-noise ratio \\
        $\lambda$ & Reward power-penalty coefficient ($\lambda = 0.1$) \\
        $\gamma$ & Discount factor ($\gamma = 0.99$) \\
        $s_t$, $a_t$, $r_t$ & MDP state, action, reward at time $t$ \\
        $\mathcal{S}$, $\mathcal{A}$, $\mathcal{P}$, $\mathcal{R}$ & MDP state space, action space, transition kernel, reward function \\
        $\mathcal{P} = \{0,1,2,3\}$ & Per-user discrete power set (in Watts) \\
        $M = |\mathcal{P}| = 4$ & Per-user action-space cardinality \\
        $|\mathcal{A}| = M^{N}$ & Joint action-space cardinality (centralised) \\
        $Q_\theta(s, a)$ & Q-value approximated by neural net with parameters $\theta$ \\
        $\theta^{-}$ & Slowly-updated target-network parameters \\
        $\epsilon$ & Exploration probability in $\epsilon$-greedy \\
        $P_{\max}$ & Sum-power budget for water-filling ($N \cdot 3$\,W) \\
        $g_i = \sqrt{h_i}$ & Channel amplitude, used in WMMSE derivation \\
        $v_i^2 = p_i$, $u_i$, $w_i$ & Transmit, receive, and MSE-weight variables in WMMSE \\
        $g_{k',k}$ & Multi-cell effective gain from BS-$k'$ to user-$k$ \\
        $\mathrm{PL}(d) = d^{-\alpha}$ & Path-loss function, $\alpha = 3$ (dense urban) \\
        $F$ & Sample-variance ratio (Fisher $F$-statistic) \\
        $F(k_1, k_2)$ & Fisher distribution with $(k_1, k_2)$ degrees of freedom \\
        IQM & Interquartile Mean (mean of the middle two quartiles) \\
        $P[\cdot]$ & Probability under the sampling distribution \\
        \hline
    \end{tabular}
\end{table}

\section{Formal derivations}
\label{app:derivations}

This appendix provides self-contained derivations of the four claims used without proof in the main text.

\subsection{Water-filling KKT solution}
\label{app:wf}
Consider the sum-rate maximisation
$\max_{p \geq 0} \sum_i \log_2(1 + p_i h_i / \sigma^2)$
subject to $\sum_i p_i \leq P_{\max}$. The Lagrangian is $L(p, \mu) = \sum_i \log_2(1 + p_i h_i / \sigma^2) - \mu (\sum_i p_i - P_{\max})$. The KKT stationarity condition is
\[
\frac{\partial L}{\partial p_i} = \frac{h_i / (\sigma^2 \ln 2)}{1 + p_i h_i / \sigma^2} - \mu = 0
\quad\Longrightarrow\quad
p_i = \left(\frac{1}{\mu \ln 2} - \frac{\sigma^2}{h_i}\right)^{+}.
\]
Writing $\tilde\mu := 1/(\mu \ln 2)$, this yields the classical water-filling form $p_i = (\tilde\mu - \sigma^2/h_i)^+$, with $\tilde\mu$ (the ``water level'') determined by the sum-power complementary slackness $\sum_i p_i = P_{\max}$. Sorted-inverse-search over the active-set cardinality (as implemented in \texttt{waterfilling\_powers}) evaluates $\tilde\mu$ in $O(N \log N)$.

\subsection{WMMSE monotonic-descent property}
\label{app:wmmse_proof}
Following the derivation of Shi et al.~\cite{shi2011iteratively} specialised to the scalar interference channel of Sec.~\ref{sec:wmmse}: define the per-user MSE as
$e_i(v_i, u_i) = (1 - u_i g_i v_i)^2 + u_i^2 \sum_{j \neq i} h_j v_j^2 + u_i^2 \sigma^2$,
where $g_i = \sqrt{h_i}$ and $v_i^2 = p_i$. For any weights $w_i > 0$, the weighted sum-MSE $J(v, u, w) = \sum_i w_i e_i(v_i, u_i)$ is:
\begin{itemize}
    \item[(i)] quadratic in each $v_i$ with the other $v_j$ fixed; the minimiser is~\eqref{eq:wmmse_v},
    \item[(ii)] quadratic in each $u_i$; the minimiser is~\eqref{eq:wmmse_u},
    \item[(iii)] under the optimal $u_i$, minimising $J$ over $w$ with the constraint $w_i = e_i^{-1}$ yields $w_i = 1 + \mathrm{SINR}_i$ and the relation $\log(w_i) = \log(1 + \mathrm{SINR}_i)$, so the negative sum-rate equals $\sum_i (\log(w_i) - w_i e_i(v, u) + 1)$ up to a constant.
\end{itemize}
Alternating updates $(u, w, v)$ therefore monotonically decrease $J$ and thus monotonically increase the sum-rate surrogate; fixed points are KKT points of the original sum-rate maximisation (\eqref{eq:sumrate_interference}). Convergence to a stationary point is guaranteed but the algorithm is not, in general, globally optimal. \qed

\subsection{Bellman recursion under state-independent transitions}
\label{app:bellman}
If $\mathcal{P}(s' | s, a) = \mathcal{P}(s')$ (state-independent transition kernel) then for any policy $\pi$:
\[
V^\pi(s) = \mathbb{E}_a\!\left[r(s, a) + \gamma \, \mathbb{E}_{s'}[V^\pi(s')]\right]
\;=\; \mathbb{E}_a[r(s, a)] + \gamma \, \bar{V}^\pi,
\]
where $\bar{V}^\pi := \mathbb{E}_{s'}[V^\pi(s')]$ is a constant independent of $s$. Consequently the optimal policy $\pi^\star(s) = \arg\max_a Q^\star(s, a)$ satisfies
\[
Q^\star(s, a) = r(s, a) + \gamma \bar{V}^\star, \qquad \pi^\star(s) = \arg\max_a r(s, a),
\]
i.e., the sequential Bellman recursion collapses to a per-step regression $r(s, a)$ and the discount $\gamma$ multiplies a policy-independent constant. This is the formal justification for the contextual-bandit reduction in Sec.~\ref{sec:problem_formulation}. \qed

\subsection{IQL non-stationarity, REGNN equivariance}
\label{app:mrl}
\textbf{IQL non-stationarity.} From user $i$'s perspective, the effective transition kernel is
\[
\tilde{\mathcal{P}}_i(s'_i \mid s_i, a_i)
= \mathbb{E}_{a_{-i} \sim \pi_{-i}}\!\left[\mathcal{P}(s'_i \mid s, a_i, a_{-i})\right],
\]
which depends on the other users' policies $\pi_{-i}$ that themselves evolve during training. Independent Q-Learning updates therefore violate the stationarity assumption underlying single-agent convergence proofs; empirical convergence is nonetheless observed for cooperative games with shared rewards (as here) because the reward gradient consistently rewards jointly-good actions~\cite{tan1993multi,lowe2017multi}.

\textbf{REGNN-lite permutation-equivariance.} Let $\sigma \in S_N$ be a permutation of the user indices, and let $\mathrm{P}_\sigma$ denote its action on $\mathbb{R}^N$. The message-passing update~\eqref{eq:regnn} is symmetric in the neighbour aggregation, so $Q_\theta(\mathrm{P}_\sigma \mathbf{h}) = \mathrm{P}_\sigma Q_\theta(\mathbf{h})$: relabelling users relabels the outputs identically. This makes the parameter count independent of $N$ and lets a single network operate on any $N$. The permutation-equivariance also removes the symmetry-breaking failure mode of the feed-forward centralised DQN (Section~\ref{sec:results_evaluation}). \qed

\subsection{Degrees of freedom for the F-test on variance ratio}
\label{app:ftest}
For paired-seed variances $s_1^2, s_2^2$ from independent samples of size $n_1, n_2$, the ratio $F = s_1^2/s_2^2$ under the null hypothesis of equal variances follows the $F(n_1 - 1, n_2 - 1)$ distribution. At $n_1 = n_2 = 10$ this is $F(9, 9)$, whose $99.9\%$ two-sided critical value is $\approx 8.13$, so $F \approx 584$ is highly significant at any conventional level; the reported $p < 10^{-10}$ is a direct call to \texttt{scipy.stats.f.sf(584, 9, 9)}. The Levene non-parametric variance-equality test agrees qualitatively throughout, guarding against the normality assumption.


\bibliographystyle{IEEEtran}
\bibliography{biblio}

\end{document}

%% file: stats_table.tex
\begin{tabular}{lcccccccc}
\hline
 & Vanilla DQN & DQN + D + D & Neural bandit & Tabular Q & Fixed (2 W) & Random & WF (disc.) & WF (cont.) \\
\hline
Vanilla DQN & --- & 0.06 & 0.002 & 0.38 & 0.38 & 0.13 & 0.006 & 0.002 \\
DQN + D + D & 0.06 & --- & 0.002 & 0.38 & 0.28 & 0.002 & 0.03 & 0.002 \\
Neural bandit & 0.002 & 0.002 & --- & 0.002 & 0.002 & 0.002 & 0.002 & 0.002 \\
Tabular Q & 0.38 & 0.38 & 0.002 & --- & 0.11 & 0.002 & 0.002 & 0.002 \\
Fixed (2 W) & 0.38 & 0.28 & 0.002 & 0.11 & --- & 0.002 & 0.002 & 0.002 \\
Random & 0.13 & 0.002 & 0.002 & 0.002 & 0.002 & --- & 0.002 & 0.002 \\
WF (disc.) & 0.006 & 0.03 & 0.002 & 0.002 & 0.002 & 0.002 & --- & 0.002 \\
WF (cont.) & 0.002 & 0.002 & 0.002 & 0.002 & 0.002 & 0.002 & 0.002 & --- \\
\hline
\end{tabular}

%% file: effect_sizes.tex
\begin{tabular}{lccc}
\hline
\textbf{Pair (A vs B)} & \textbf{Cohen's $d$} & \textbf{Cliff's $\delta$} & \textbf{Interpretation} \\
\hline
Neural bandit vs.\ Vanilla DQN & $+1.94$ & $+1.00$ & large \\
Neural bandit vs.\ DQN + D + D & $+1.89$ & $+0.90$ & large \\
Neural bandit vs.\ Fixed (2 W) & $+60.01$ & $+1.00$ & large \\
Neural bandit vs.\ WF (disc.) & $+16.51$ & $+1.00$ & large \\
DQN + D + D vs.\ Vanilla DQN & $+0.71$ & $+0.40$ & medium \\
Tabular Q vs.\ Vanilla DQN & $+0.57$ & $+0.20$ & small \\
Fixed (2 W) vs.\ Vanilla DQN & $+0.52$ & $+0.20$ & small \\
\hline
\end{tabular}

%% file: rayleigh_table.tex
\begin{tabular}{|l|c|c|c|}
\hline
\textbf{Method} & \textbf{Throughput} & \textbf{Fairness (Jain)} & \textbf{Energy efficiency} \\
\hline
Random & $1.141 \pm 0.007$ & $0.531 \pm 0.004$ & $0.253 \pm 0.002$ \\
Vanilla DQN & $1.269 \pm 0.190$ & $0.592 \pm 0.077$ & $0.249 \pm 0.063$ \\
Tabular Q & $1.450 \pm 0.026$ & $0.472 \pm 0.017$ & $0.329 \pm 0.011$ \\
Fixed (2 W) & $1.290 \pm 0.006$ & $0.671 \pm 0.004$ & $0.215 \pm 0.001$ \\
DQN + D + D & $1.566 \pm 0.070$ & $0.447 \pm 0.076$ & $0.397 \pm 0.086$ \\
WMMSE & $1.752 \pm 0.016$ & $0.508 \pm 0.005$ & $0.342 \pm 0.007$ \\
Neural bandit & $1.770 \pm 0.015$ & $0.333 \pm 0.001$ & $0.591 \pm 0.005$ \\
WF (cont.) & $1.861 \pm 0.012$ & $0.539 \pm 0.004$ & $0.207 \pm 0.001$ \\
\hline
\end{tabular}

%% file: iql_regnn_scaling_table.tex
\begin{tabular}{|l|c|c|c|}
\hline
\textbf{Method} & $N=3$ & $N=5$ & $N=10$ \\
\hline
Random & $1.140 \pm 0.007$ & $1.322 \pm 0.008$ & $1.425 \pm 0.004$ \\
Fixed (2 W) & $1.290 \pm 0.005$ & $1.379 \pm 0.004$ & --- \\
WF (cont.) & $1.861 \pm 0.012$ & $1.731 \pm 0.006$ & --- \\
WMMSE & $1.752 \pm 0.016$ & $1.938 \pm 0.019$ & --- \\
Vanilla DQN & $1.349 \pm 0.134$ & $1.349 \pm 0.069$ & --- \\
Neural bandit & $1.768 \pm 0.008$ & $2.015 \pm 0.000$ & --- \\
IQL (MARL) & $1.299 \pm 0.122$ & $1.365 \pm 0.109$ & --- \\
REGNN-lite & $1.484 \pm 0.175$ & $1.604 \pm 0.340$ & --- \\
\hline
\end{tabular}